\documentclass[12pt,preprint]{aastex}

\usepackage{psfig}

\slugcomment{}

\shorttitle{Dark Matter ``Temperature'' in Abell 1795}
\shortauthors{Ikebe et al.}

\begin{document}


\title{X-ray Measurement of Dark Matter ``Temperature'' in Abell 1795}


\author{Yasushi Ikebe\altaffilmark{1}}
\affil{Joint Center for Astrophysics, University of Maryland, Baltimore County,
	1000 Hilltop Circle, Baltimore, MD 21250, USA}
\email{ikebe@milkyway.gsfc.nasa.gov}

\author{Hans B\"{o}hringer}
\affil{Max-Planck-Institut f\"{u}r extraterrestrische Physik,
	Postfach 1312, 85741 Garching, Germany}

\and

\author{Tetsu Kitayama}
\affil{Department of Physics, Toho University,
	Miyama, Funabashi, Chiba 274-8510, Japan}


\altaffiltext{1}{Office address: Code 661, NASA/Goddard Space Flight Center, Greenbelt Rd., Greenbelt, MD 20771, USA}


\begin{abstract}
We present a method from an X-ray observation of a galaxy cluster
to measure the radial profile of the dark matter velocity dispersion,
$\sigma_{\rm DM}$, and to compare the dark matter ``temperature''
defined as $\mu m_{\rm p} \sigma_{\rm DM}^2 / k_{\rm B}$
with the gas temperature.
The method is applied to the XMM-Newton observation of Abell 1795.
The ratio between the specific energy of the dark matter
and that of the intracluster medium (ICM),
which can be defined as $\beta_{\rm DM}$ in analogy with $\beta_{\rm spec}$,
is found to be less than unity everywhere ranging $\sim 0.3-0.8$.
In other words,
the ICM temperature is higher than the dark matter ``temperature'',
even in the central region where the radiative cooling time is short.
A $\beta_{\rm DM}$ value smaller than unity can most naturally be explained
by heating of the ICM.
The excess energy of ICM is estimated to be $\sim1-3$~keV per particle.
\end{abstract}




\section{Introduction}
Early X-ray imaging observations with the {\it Einstein} observatory
and {\it ROSAT} showed that
in the central regions of clusters of galaxies
the radiative cooling time is shorter than the age of the universe
(e.g Canizares, Stewart, \& Fabian 1983).
As a result, the intracluster medium (ICM) should cool down to form
a cold ($T<10^6$K) gas phase inducing a global inflow of gas.
This ``cooling flow'' (see Fabian 1994 for a review) picture
has been extensively discussed
and formed a basic assumption in many arguments.
The low resolution spectroscopy in 0.5-2 keV by {\it ROSAT} showed
that in some clusters
the ICM temperature actually decreases towards the center
(e.g. B\"{o}hringer et al. 1994; David et al. 1994;
Allen \& Fabian 1994).
Higher resolution spectroscopy in 0.5-10 keV with {\it ASCA},
however, can not be fully understood with the conventional cooling flow model.
{\it ASCA} spectra of cooling flow clusters 
can be well explained by a two (hot and cool)
phase plasma without significant excess absorption features
(e.g. Ikebe et al. 1999; Makishima et al. 2001).
A naive cooling flow model predicting a range of temperatures with intrinsic
absorption could also fit the {\it ASCA} data but generally produce
worse chi-square results (e.g. Allen et al. 2001).
Most recently, very high resolution spectroscopy with XMM-Newton/RGS
unambiguously show that there is very little X-ray emission from
gas cooler than certain lower cut-off temperatures of $\sim 1-3$~keV
(Tamura et al. 2001; Kaastra et al. 2001; Peterson et al. 2001).
Unless a large amount of cooled gas 
or the metals in the cold gas are hidden (Fabian et al. 2001a),
there must exist a heating mechanism that prevents
the ICM from radiative cooling.

The necessity of global heat input into the ICM
in addition to gravitational heating has also been pointed out
from the break of the self-similarity between dark matter and ICM,
which is most clearly demonstrated in the X-ray luminosity-temperature
relation ($L-T$ relation).
A simple scaling analysis (Kaiser 1986) suggests a relation of $L\propto T^2$,
while observation shows $L\propto T^{3}$
(e.g. Edge \& Stewart 1991; David et al. 1993; White et al. 1997;
Wu et al. 1999).
The break of the self-similarity is also seen
in the entropy vs temperature relation.
Ponman, Cannon, \& Navarro (1999) showed 
that cooler systems ($T<4$ keV) have entropies higher than
achievable through gravitational collapse alone.
Simulation studies show that the observed relations can be reproduced,
if there is enough non-gravitational heat input into the ICM
by feedback from galaxies
(e.g. Metzler \& Evrard 1994; Bower et al. 2001)
or preheating before the cluster formation 
(e.g. Navarro et al. 1995; Tozzi \& Norman 2000).

In order to shed some new light onto these ``cooling flow phenomena''
and ``break of self similarity'',
we, in the present paper,
perform a comparison of the temperature distribution of the ICM
to the distribution of the velocity dispersion of the dark matter.
A parameter, $\beta_{\rm spec}\equiv\sigma_{gal}^2/(k_{\rm B}T/\mu m_p)$,
is often used as a measure of the average kinetic energy per unit mass
in galaxies relative to that in the ICM.
From observations of many clusters,
the mean $\beta_{\rm spec}$ is $\sim 1$ with large scatter
(e.g. Wu et al. 1999),
indicating that the energy equipartition between galaxies
and ICM is roughly achieved on average.
In analogy with $\beta_{\rm spec}$,
we can introduce $\beta_{\rm DM}\equiv\sigma_{DM}^2/(k_{\rm B}T/\mu m_p)$
for comparison between the mean kinetic energy of the dark matter
and that of the ICM, and define the dark matter ``temperature''
as $T_{\rm DM}\equiv\mu m_{\rm p} \sigma_{\rm DM}^2 / k_{\rm B}$.
Note that the above definition of ``temperature'',
by means of proton mass instead of the actual mass of dark matter
particles, is only for the sake of comparison with the gas temperature.
Therefore we put ``temperature'' in quotes.
We obtain, in this paper, the radial profile of the $\beta_{\rm DM}$ value
observationally for the first time.


In Sect. 2, we describe the method of measuring
the dark matter velocity dispersion in a cluster of galaxies
from an X-ray observation.
We applied the method to the XMM-Newton data
of a prototypical cooling flow cluster,
Abell~1795 (hereafter A1795),
which is located at z=0.0616 (Struble \& Rood 1987).
The X-ray data analysis and results are presented in Sect. 3.
Discussion and summary are found in Sect. 4 and Sect. 5,
respectively.
Throughout the paper, the Hubble constant is given
as 70 $h_{70}$ km s$^{-1}$ Mpc$^{-1}$,
and a flat universe 
($\Omega_{\rm m,0}=0.3$, $\Omega_{\rm \Lambda,0}=0.7$) is assumed.
At the redshift of A1795, 1 arcsec corresponds to 1.19 kpc.

\section{Method of measuring dark matter velocity dispersion}

We use a simplified model cluster being composed of a hot plasma ICM 
with a temperature of $10^7-10^8$~K,
and dark matter made of collisionless particles.
Under the assumptions of spherical symmetry and hydrostatic equilibrium,
the ICM distribution is described by
\begin{equation}
\frac{GM}{R} = - \frac{k_{\rm B}T_{\rm g}}{\mu m_p}
\left(\frac{d\ln{n_{\rm g}}}{d \ln{R}} + \frac{d\ln{T_{\rm g}}}{d\ln{R}}\right)\ ,
\label{eq:hydro}
\end{equation}
where $M(<R)$ is the total gravitating mass within a sphere of radius $R$,
$k_{\rm B}$ is the Boltzmann constant, 
$G$ is the Gravitational constant,
$n_{\rm g}(R)$ and $T_{\rm g}(R)$ is the density and temperature of the ICM,
respectively.
From an X-ray observation, $n_{\rm g}(R)$ and $T_{\rm g}(R)$ are measured
and $M(<R)$ can be obtained via Eq.~(\ref{eq:hydro}).
When the dark matter particles tracing the same gravitational field
are in steady state,
they obey the Jeans equation
\begin{equation}
\frac{GM}{R} = 
- \sigma_{\rm DM}^2 \left( \frac{d \ln{\rho_{\rm DM}}}{d \ln{R}} 
 + \frac{d \ln{\sigma_{\rm DM}^2}}{d \ln{R}} \right) \ ,
\label{eq:Jeans}
\end{equation}
where $\sigma_{\rm DM}(R)$ is the one-dimensional radial velocity dispersion,
and $\rho_{\rm DM}$ is the mass density, which is given as
$\rho_{\rm DM} = \frac{1}{4\pi R^2}\frac{dM}{dR} - \mu m_{\rm p} n_{\rm g}$.
Therefore, once the total gravitating mass, $M$, and the gas density,
$n_{\rm g}$, profiles are obtained from an X-ray observation,
Eq.~(\ref{eq:Jeans}) contains only one unknown parameter, $\sigma_{\rm DM}(R)$.
Equation~(\ref{eq:Jeans}) can be rewritten in the form of 
an ordinary differential equation for $\sigma_{\rm DM}^2$ as
\begin{equation}
\frac{d\sigma_{\rm DM}^2}{dR} 
= - \frac{GM}{R^2} 
  - \frac{\sigma_{\rm DM}^2}{\rho_{\rm DM}}\frac{d\rho_{\rm DM}}{dR}\ ,
\label{eq:diffeq}
\end{equation}
which can be solved numerically under a given boundary condition
to derive the velocity dispersion profile of the dark matter.

Although a steady state dark matter distribution is assumed above,
a dark matter density profile should actually be growing
as matter is falling onto the system from outside.
A numerical simulation by Fukushige \& Makino (2001) shows
that the dark matter halo grows in a self-similar way,
keeping the density profile in the central region unchanged.
This justifies the assumption at least in the central region.

A similar technique as described above
was used in determining the ICM temperature profile,
using X-ray imaging data without spectroscopic information
taken e.g. with the {\it Einstein} observatory.
A model profile for the total mass is assumed 
and the temperature profile is derived from the equation of hydrostatic
equilibrium so that the observed brightness profile is explained 
(e.g. Fabian et al. 1981; Hughes 1989).
Here we solve the Jeans equation instead in order to derive
the dark matter velocity dispersion.

\section{A1795 and the XMM-Newton observation}

An ideal opportunity to study the mass profile as well as velocity dispersion
profile of dark matter is provided by the XMM-Newton observation of
a prototypical cooling flow cluster, A1795.
Early X-ray imaging observations with the {\it Einstein} observatory
and {\it ROSAT} showed that
the X-ray emission from A1795 has almost circular symmetry
with a small elongation along the north-south direction,
which indicates that the cluster is well relaxed dynamically
(Jones \& Forman 1984; Boute \& Tsai 1996).
The radial brightness profile shows a huge central excess above
a prediction from an isothermal $\beta$-model profile,
and the central excess luminosity gives a mass deposition rate
of $\sim250h_{70}^{-2}$ M$_{\odot}$ yr$^{-1}$
based on the standard cooling flow interpretation
(Edge et al. 1992; Briel \& Henry 1996).
From the 0.5-10 keV spectrum taken with {\it ASCA}, however,
a significantly smaller mass deposition rate is obtained
($\sim66h_{70}^{-2}$ M$_{\odot}$ yr$^{-1}$ Fabian et al. 1994;
$\sim70h_{70}^{-2}$ M$_{\odot}$ yr$^{-1}$ Xu et al. 1998;
$\sim150h_{70}^{-2}$ M$_{\odot}$ yr$^{-1}$ Allen et al. 2001).
Instead of the cooling flow model,
the {\it ASCA} spectrum can be best described with a two temperature model
without excess absorption (Xu et al. 1998).

The XMM-Newton observation of A1795 was carried out during
the performance-verification phase on June 26 2000 with
total observing time of $\sim50$ ksec.
The ICM temperature profile obtained with EPIC MOS as well as EPIC PN
were already reported by Tamura et al. (2001) and by Arnaud et al. (2001),
showing that it is almost isothermal at $\sim$6 keV
in the 2-10 arcmin radius region,
while it decreases towards the center reaching the minimum temperature
at $\sim3$ keV.
The Chandra observation gives consistent results (Ettori et al. 2002).
A high resolution spectrum from the central region obtained with
XMM-Newton/RGS has been analyzed by Tamura et al. (2001),
showing a lack of signature of cool gas components below $\sim3$ keV
and an upper limit of 77 $h_{70}^{-2}$ M$_{\odot}$ yr$^{-1}$
for the mass deposition rate 
is obtained if an isobaric cooling flow model is applied.

In this paper, we analyze the data taken with
the XMM-Newton/EPIC PN, MOS1 and MOS2
to measure the total mass profile of A1795 
and to derive the dark matter velocity dispersion profile
using the method described in Sect. 2.

\section{Analysis of the XMM-Newton data and results}

\subsection{Data screening and background subtraction}

A significant fraction of any XMM-Newton observation is often
contaminated by the huge background count rate by soft proton flares.
In order to derive time intervals with stable background,
we made the 0.5--10 keV band lightcurves individually with
the PN, MOS1, and MOS2 data,
and eliminated time periods where the count rate deviates
from the mean value during quiescent periods by $\ge 2\sigma$.
The total usable exposure time thus left is 23 ksec, 31 ksec, and 34 ksec
for PN, MOS 1 and MOS 2, respectively.
In the present analysis, the single and double pixel events of the PN data
and the single, double, triple, and quadruple events of the MOS data are
used.

The background consisting of the cosmic X-ray background
and the high energy particle events was estimated from
the Lockman Hole data taken during the revolution \# 70.
In this sky region, the Galactic column density is as low as the A1795 field.
After eliminating the contamination sources from the Lockman Hole data
as well as the A1795 data, 
the Lockman Hole data are scaled so that 
the count rates of the background and the cluster data becomes the same
in the 7-12 keV energy band,
and in the radius range of 575-925 arcsec for PN
and of 500-850 arcsec for MOS,
where no significant X-ray signal from the cluster was detected.
The background data thus obtained were subtracted from the cluster data.
For PN, the out-of-time events (Str\"{u}der et al. 2001)
were also subtracted prior to the subtraction of the background.

\subsection{Radial count rate profile}

\begin{figure}
\epsscale{0.6}
\plotone{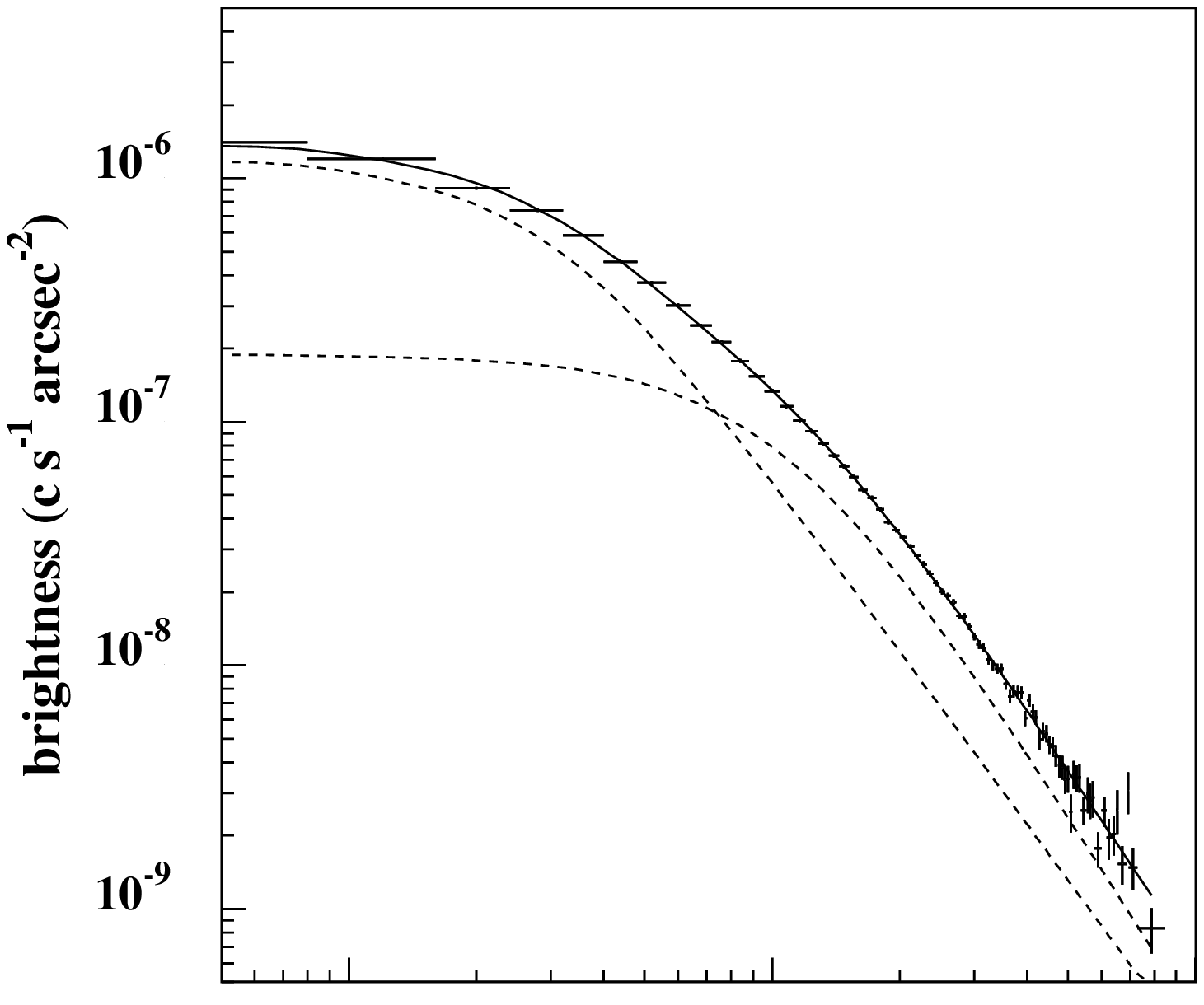}
\plotone{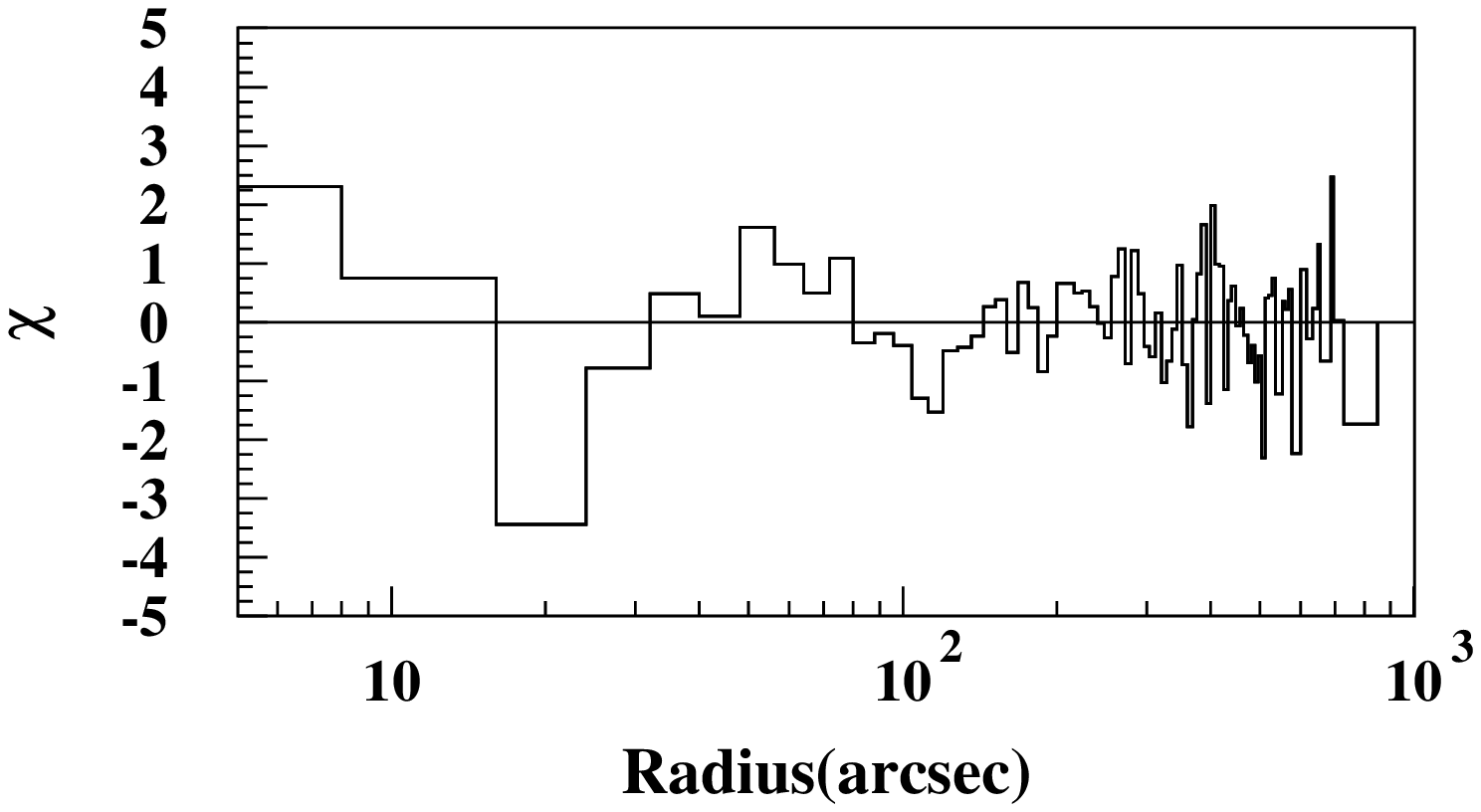}
\caption{In the upper panel, the 0.8--10~keV X-ray count rate profile
shown in crosses is fitted with a double $\beta$-model (solid line).
The dotted lines show the individual $\beta$-model components.
The fit residuals are shown in the lower panel.
\label{fit_2beta}}
\end{figure}

Defining the center as the X-ray peak position,
we have produced a projected radial brightness profile in the 0.8-10 keV band.
The X-ray count rate was converted to the X-ray flux
by correcting each X-ray photon with the instruments' responses
that include the effective area and vignetting of the X-ray telescopes,
and the quantum efficiency of the CCD cameras.
Combining the data taken with PN, MOS1 and MOS2,
we thus derived the brightness profile,
which is illustrated in Fig.~\ref{fit_2beta}.
For comparison with the other X-ray observations,
we quantified the radial profile using the $\beta$-model profile
given by
\begin{equation}
\Sigma = \Sigma_{\rm 0}
\left[ 1 + \left(\frac{R}{R_{\rm c}}\right)^2 \right]^{-3\beta+0.5}\ ,
\label{eq:beta}
\end{equation}
where $R_{\rm c}$ is a core radius and $\beta$ is a beta parameter.
The $\beta$ model has been often used to fit X-ray brightness profiles
and generally gives a good representation for non cooling flow clusters.
Since the X-ray brightness profile of A1795 has long been known
to show a central excess above a $\beta$-model profile,
we employed here a double $\beta$-model profile
that is the sum of two $\beta$-model profiles
to fit the 0.8-10 keV brightness profile within 848 arcsec.
In the actual fitting,
the model brightness profile is convolved with
the point spread function of the X-ray telescope
\footnote{
The PSF is approximated by an analytic function given as
$\left[ 1 + \left(\frac{r}{r_{\rm c}}\right)^2\right]^{-\alpha}$,
where $r_{\rm c}$ and $\alpha$ depend on off-axis angle as well as X-ray
energy (see Ghizzardi 2001).
We assumed that the PSF is constant in the entire FOV
and independent of the X-ray energy.
As the representative, we employed the on-axis PSF at 1.5 keV.
Using the $r_{\rm c}$ and $\alpha$ parameters of
($r_{\rm c}$[arcsec], $\alpha$)
= (5.37, 1.5), (4.72, 1.457), and (4.4545, 1.4035),
for PN, MOS1, and MOS2, respectively,
we obtained the average PSF,
which is then used in the brightness profile fittings.
},
whose half-power-radius is $\sim9$ arcsec at 1.5 keV on-axis
and significantly affects several of the central bins.
Evaluating the goodness of the fit with a chi-square statistic,
we obtained a good fit as shown in Fig.~\ref{fit_2beta}
with the best-fit parameters as summarized in table \ref{tab:2beta}.

\placefigure{fit_2beta}

\placetable{tab:2beta}

\subsection{ICM temperature and metallicity profile}

The temperature and metallicity profiles given by Tamura et al. (2001)
were derived from the conventional annular spectral analysis,
accumulating spectra from concentric annular regions
in the projected 2-dimensional space.
We derive here the temperature and metallicity profiles
instead in 3-dimensional form from deprojected spectra.
With each PN, MOS1, and MOS2 data set,
we first accumulate 13 annular spectra
from the vignetting corrected count rate profiles
in individual energy channels.
The outer radii are
16, 32, 48, 64, 96, 128, 192, 256, 384, 512, 640, 768, and 848 arcsec.
Assuming spherical symmetry and that
there is no cluster emission beyond 848 arcsec
from the cluster center,
we deprojected the annular spectra to spectra
for 13 spherical shell regions whose radii correspond to the annular radii.
Since no intrinsic absorption is found even from the central region
(Tamura et al. 2001),
the X-ray emission from the cluster is entirely optically thin,
and we performed the deprojection process as follows.
The relation between
the energy spectra observed in the $N$ annular regions,
$S_{\rm j}(E)$ ($j = 1 - N$), and those emitted from the $N$ shell regions,
$C_{\rm i}(E)$ ($i = 1 - N$),
is given with a matrix relation as
\begin{equation}
S_{\rm j}(E) = \sum_{i=1}^{N} M_{\rm j, i} C_{\rm i}(E) \ ,
\end{equation}
where $N=13$ in the present case.
The matrix elements are given as
\begin{equation}
M_{\rm j,i} 
 = \left\{
 \begin{array}{ll}
 [ (r_{\rm i+1}^2 - r_{\rm j}^2)^{3/2} - (r_{\rm i}^2 - r_{\rm j}^2)^{3/2}
 - (r_{\rm i+1}^2 - r_{\rm j+1}^2)^{3/2} + (r_{\rm i}^2 - r_{\rm j+1}^2)^{3/2} ]
 / (r_{\rm i+1}^3 - r_{\rm i}^3)
   &   {\rm if}\ i>j,\\
 (r_{\rm i+1}^2 - r_{\rm i}^2)^{3/2} / (r_{\rm i+1}^3 - r_{\rm i}^3) & {\rm if}\ i=j,\\
   0   &   {\rm if}\ i<j,
 \end{array}
 \right.
\end{equation}
where $r_{\rm i}$ ($r_{\rm j}$) and $r_{\rm i+1}$ ($r_{\rm j+1}$)
are the inner and the outer radius of the i'th shell (j'th annular),
respectively.
Each element represents the volume fraction of the i'th shell contributing
to the emission detected in the j'th annular.
Since $M_{\rm j,i}$ is a triangular matrix,
it can be easily inverted to $M_{\rm i,j}^{-1}$,
and the deprojected shell region spectra
can be derived from the annular spectra by
\begin{equation}
C_{\rm i}(E) = \sum_{j=1}^{N} M_{\rm i, j}^{-1} S_{\rm j}(E) \ .
\end{equation}
Statistical errors in the individual energy bins are properly
propagated through this operation.

Out of the 13 deprojected spectra thus derived from each PN, MOS1,
and MOS2 data set,
we used only the inner 10 deprojected spectra
within a 512 arcsec radius in the analysis below
so that
the artificial cut-off of the X-ray emission outside the maximum radius
of 848 arcsec does not affect 
the resulting deprojected spectra within 512 arcsec.
In addition, the signal to noise ratios of the spectra within 512 arcsec
are high enough,
and some 10\% systematic error of the predicted background count rates
introduce no significant systematic errors on temperature measurements.

\begin{figure}
\epsscale{0.8}
\plotone{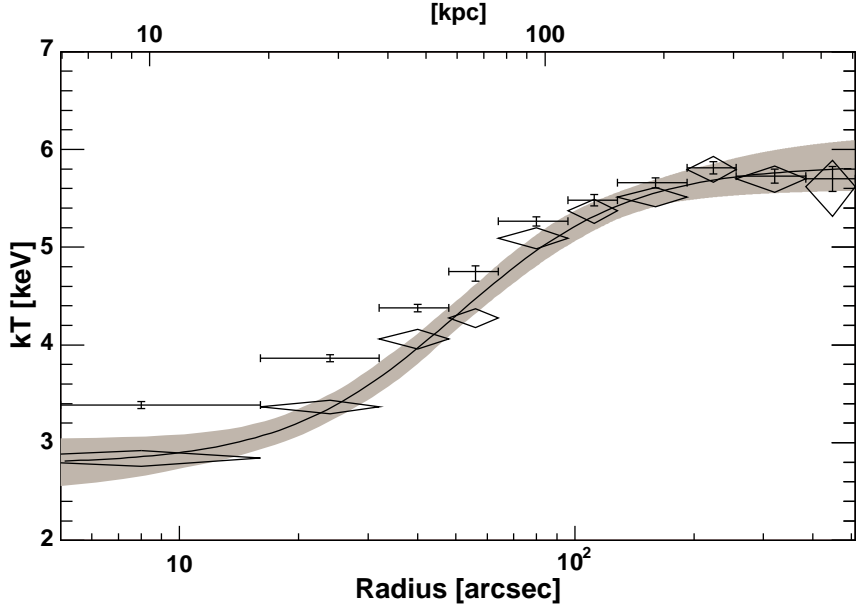}
\caption{
The ICM temperature profile of A1795 derived
from the spectral analysis with the XMM-Newton EPIC-MOS
and EPIC-PN data.
Results from the deprojected spectra are shown with diamonds,
while the conventional annular spectral analysis gives 
the profile with crosses.
The solid line shows the best-fit function given as
$ T(R)=2.78+3.06 \left[ 1 + \left(\frac{R(arcsec)}{50.3}\right)^{-1.98} \right]^{-1} $ (keV),
while the gray hatched region shows 90\% error region with the function.
\label{tpro}}
\end{figure}

\begin{figure}
\epsscale{0.8}
\plotone{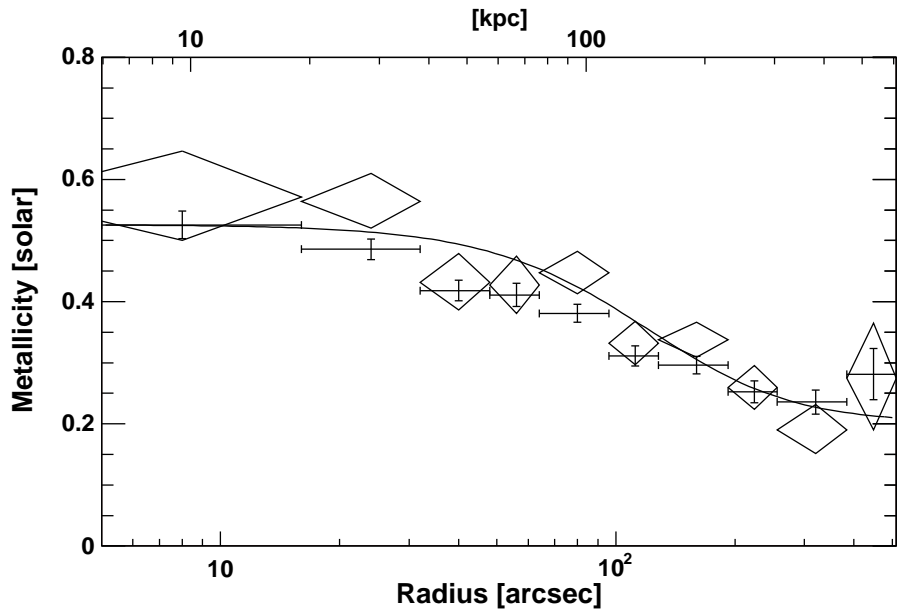}
\caption{The metallicity profile of A1795.
Results from the deprojected spectra are shown with diamonds,
while the conventional annular spectral analysis
gives the profile with crosses.
The solid line shows the best-fit function given as
$A(R)=0.20+0.32 (1+(R/149)^2)^{-1.5}$.
\label{apro}}
\end{figure}

In order to derive the temperature and metallicity profile,
we simultaneously fitted the PN, MOS1, and MOS2 spectra
with a single-temperature plasma model in each deprojected shell region.
We employed the MEKAL model 
(Mewe et al. 1985, 1986; Kaastra 1992; Liedahl et al. 1995)
and assumed
the abundance ratio among different elements to have solar values.
The fits are acceptable for all the ten deprojected shell regions.
The deprojected temperature and metallicity profiles thus derived
are illustrated in Fig.~\ref{tpro} and Fig.~\ref{apro}, respectively.
Compared with the result from the conventional annular spectral analysis,
the effect of the projection is clearly seen inside the $\sim$60 arcsec radius.

The temperature is consistent with being isothermal
in the outskirts beyond $\sim$200~kpc,
while towards the center it decreases monotonically 
and reaches 2.8~keV in the innermost shell.
Note that the central minimum temperature agrees with 
the cut-off temperature in a cooling flow model
that describes the RGS spectrum (Tamura et al. 2001)
and the temperature of the X-ray filament detected with Chandra
(Fabian et al. 2001).
For usage in the following sections,
we obtained an analytical formula that approximates the temperature profile.
We employed a formula given as
\begin{equation}
T(R) = T_0 + T_1 \left[ 1 + \left(\frac{R}{R_{\rm c,T}}\right)^{-\eta} \right]^{-1} \,
\label{eq:tpro}
\end{equation}
which has been proposed by Allen, Schmidt, and Fabian (2001)
to give a good description to radial temperature profiles in relaxed clusters.
Leaving all the four parameters of $T_0$, $T_1$, $R_{\rm c,T}$, and $\eta$
as free parameters, 
we fitted the temperature profile derived from the deprojected spectra.
The parameters obtained from a $\chi^2$ minimization are
$T_0=2.78^{+0.28}_{-0.38}$ (keV), $T_1=3.06^{+0.68}_{-0.45}$ (keV),
$R_{\rm c,T}=50.3^{+10.9}_{-8.6}$ (arcsec), and $\eta=1.98^{+0.82}_{-0.65}$,
where the errors give the 90\% confidence range
for four parameters of interest ($\Delta \chi^2 = 7.78$).
The best-fit function and the error ranges for the individual radii
are overlaid in Fig.~\ref{tpro}, 
showing that the function gives a good representation
of the temperature profile.

\placefigure{tpro}

\placefigure{apro}

In the metallicity profile (Fig.~\ref{apro}),
the strong central concentration of metals 
found in the previous works (Tamura et al. 2001; Ettori et al. 2002) is confirmed.
As for the temperature profile,
the metallicity profile was also fitted with a similar function given as
\begin{equation}
A(R) = A_0 + A_1 \left[ 1 + \left(\frac{R}{R_{\rm c,A}}\right)^2 \right]^{-3/2} \ .
\label{eq:apro}
\end{equation}
We derived $A_0=0.20^{+0.10}_{-0.19}$ (solar),
$A_1=0.32^{+0.15}_{-0.11}$ (solar),
and $R_{\rm c,A}=149^{+170}_{-82}$ (arcsec),
and the best-fit function is overlaid in Fig.~\ref{apro}.

\subsection{Mass profile}

An immediate way to derive the mass profile is using Eq. (\ref{eq:hydro})
with the temperature profile measured in Sect. 4.3
and the density profiles of the ICM which can be obtained from
the double $\beta$-model profile fitted to the observed X-ray brightness
profile in Sect. 4.2.
Assuming spherical symmetry,
the projected count rate profile in the 0.8-10 keV band,
$\Sigma(0.8-10 {\rm keV})$,
gives the deprojected radial emissivity profile, $\epsilon(0.8-10 {\rm keV})$,
which can be converted to the ICM density via the relation
\begin{equation}
\epsilon(0.8-10 {\rm keV}) = n_{\rm g}^2 \Lambda(T,A;0.8-10 {\rm keV}) \ ,
\label{eq:n2epsilon}
\end{equation}
where $\Lambda(T,A;0.8-10 {\rm keV})$ is the emissivity coefficient
defined by temperature ($T$), metallicity ($A$), energy range (0.8-10 keV),
and the instrument responses.

\begin{figure}
\epsscale{0.8}
\plotone{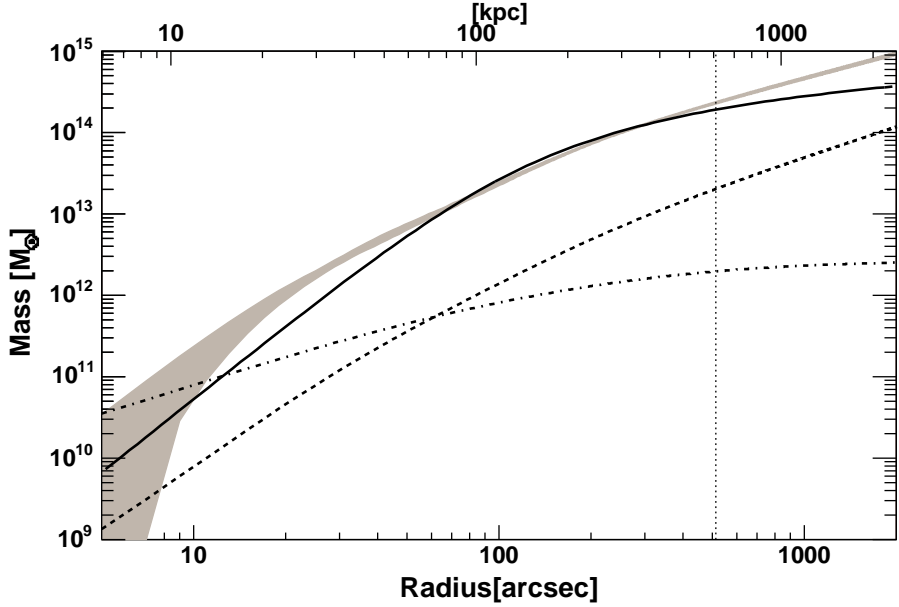}
\caption{
The grey hatched region illustrates the integrated radial profile
of the total gravitating mass obtained via Eq. (\ref{eq:hydro}).
The errors in the ICM temperature profile illustrated in
Fig.~\ref{tpro} is properly propagated to obtain the error range
at the individual radii.
The vertical dotted line indicates the radius within
which the ICM temperature measurements are available,
and isothermality is assumed beyond the radius.
The dashed line shows the ICM mass profile calculated
from integrating the ICM density profile,
where the errors are smaller than the line width.
The dot-dashed line shows the stellar mass profile,
estimated from an $I$-band image obtained by Johnstone et al. (1991).
We approximated the $I$-band surface brightness profile with
a de Vaucouleurs law given as
$\mu [{\rm mag\ arcsec^{-2}}] = 
25.0 + 8.33 \left[\left\{\frac{r({\rm arcsec})}{150}\right\}^{1/4}-1\right]$,
and converted it to the mass profile
assuming $B$--$I$=2.28 and a mass-to-light ratio of 6.5 in $B$-band.
The solid line indicates an approximated King profile
(Eq. \ref{eq:king_rho} and \ref{eq:king_mass})
derived by fitting it to the data via Eqs.~(\ref{eq:n2epsilon}) and 
(\ref{eq:m2n}).
\label{masspro}}
\end{figure}

Taking into account the error ranges in the temperature profile,
we thus derived the total gravitating mass profile,
which is shown in Fig. \ref{masspro}.
The result roughly agrees with the mass profile given by Xu et al. (1998)
derived from the {\it ASCA} data with the same method,
although a shoulder-like structure found in their profile at $\sim 100$~kpc
is less prominent in our profile.
The difference in the mass profiles is mainly attributed to
the different parameters of the double $\beta$-model profile
fitted to the brightness profile.

\placefigure{masspro}

\subsection{Theoretical modeling to the mass profile}
The total mass profile thus derived can not be conveniently
used in Eq. (\ref{eq:diffeq})
for determining the dark matter velocity dispersion, however.
The term of $d\rho_{\rm DM}/dR$ in Eq. (\ref{eq:diffeq})
is calculated from the third derivative of the ICM density profile,
i.e. the third derivative of the radial brightness profile
which is approximated with the double-$\beta$ profile,
and it gives an uneven profile at
the radius where the two $\beta$ profiles cross over.
This causes an unstable behavior in the solution of
Eq. (\ref{eq:diffeq})
and introduces non-negligible systematic errors in the results.
Moreover, the double-$\beta$ modeling on the brightness profile
can only reproduce a mass profile with the flat core 
as long as the temperature profile is virtually flat near the center,
and provides rather restrictive mass profile.

Here we make an alternative approach from the derivation above.
We start modeling the total mass profile with an analytical formula
with a few free parameters
that assures a smooth profile with Eq. (\ref{eq:diffeq}).
Combined with the temperature profile derived in Sect. 4.3,
the mass profile is used to predict the X-ray surface brightness
profile of the cluster to be compared with the data to adjust the free parameters.
In the actual calculation,
a given total mass profile, $M(<R)$, and a temperature profile, $T(R)$,
is converted to the ICM density profile from Eq. (\ref{eq:hydro})
as
\begin{equation}
n_{\rm g}(R) = n_0 \frac{T(0)}{T(R)} 
   \exp\left[ - \int_0^R \frac{G\mu m_{\rm p} M(<R)}{kT(R)R^2}dR \right]\ ,
\label{eq:m2n}
\end{equation}
where $n_0$ is the central ICM density.
The emissivity profile in a given energy range, $\epsilon(E_1,E_2)$,
is then obtained by Eq. (\ref{eq:n2epsilon}),
where $n_{\rm g}$, $T$, $A$ are substituted by Eqs. (\ref{eq:m2n}),
(\ref{eq:tpro}) and (\ref{eq:apro}), respectively,
which is then converted to the brightness profile, $\Sigma(E_1,E_2)$.
Being convolved with the PSF,
the model brightness profile is fitted
to the 0.8-10~keV count rate profile observed
to determine the best-fit parameters in the mass profile model.
Although the observational constraint on the ICM temperature profile
is limited within 512 arcsec from the center,
we apply Eq. (\ref{eq:tpro}) beyond the radius,
where it describes a practically isothermal model.

\begin{figure}
\centerline{
\psfig{file=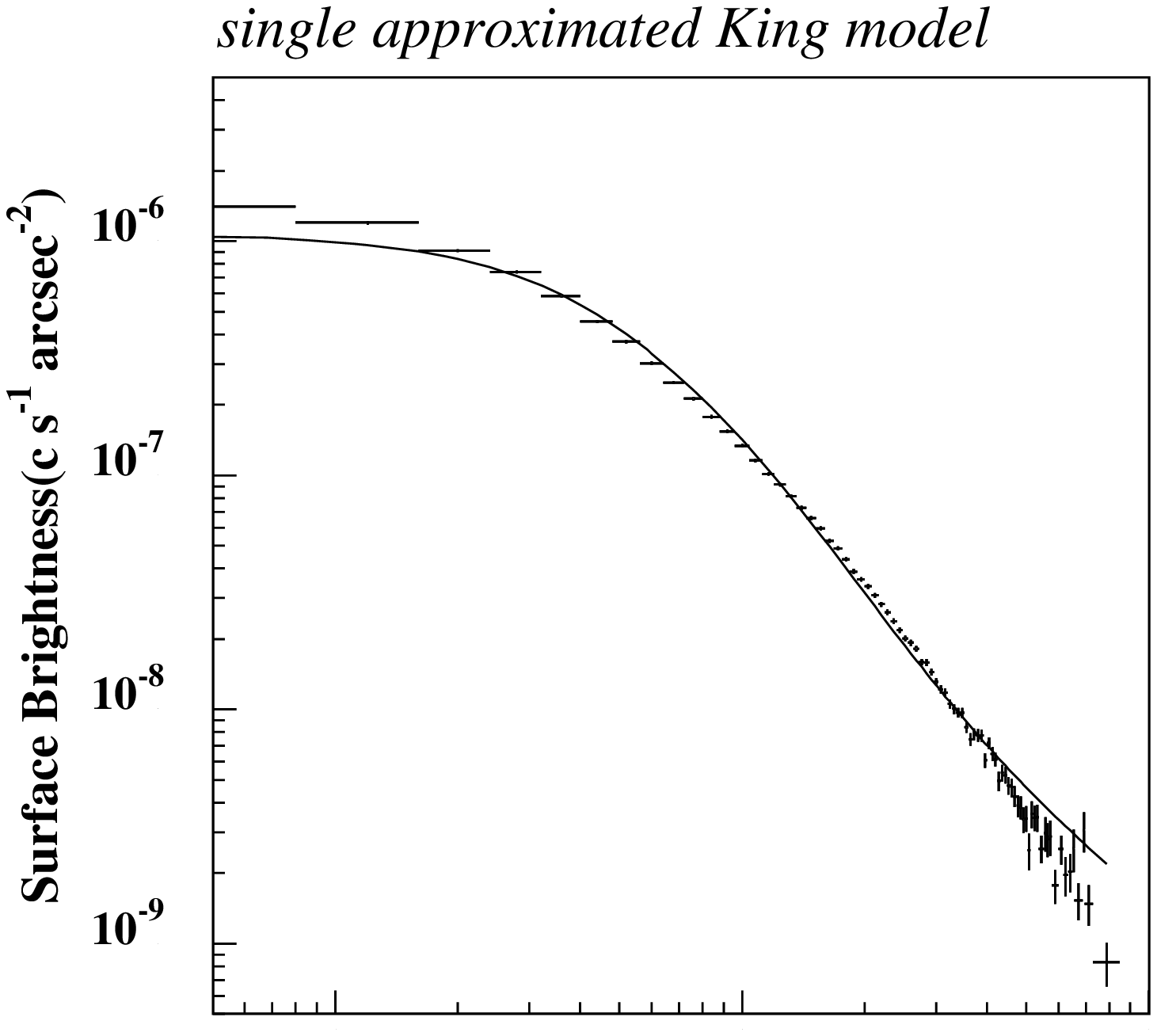,width=5.5cm}
\psfig{file=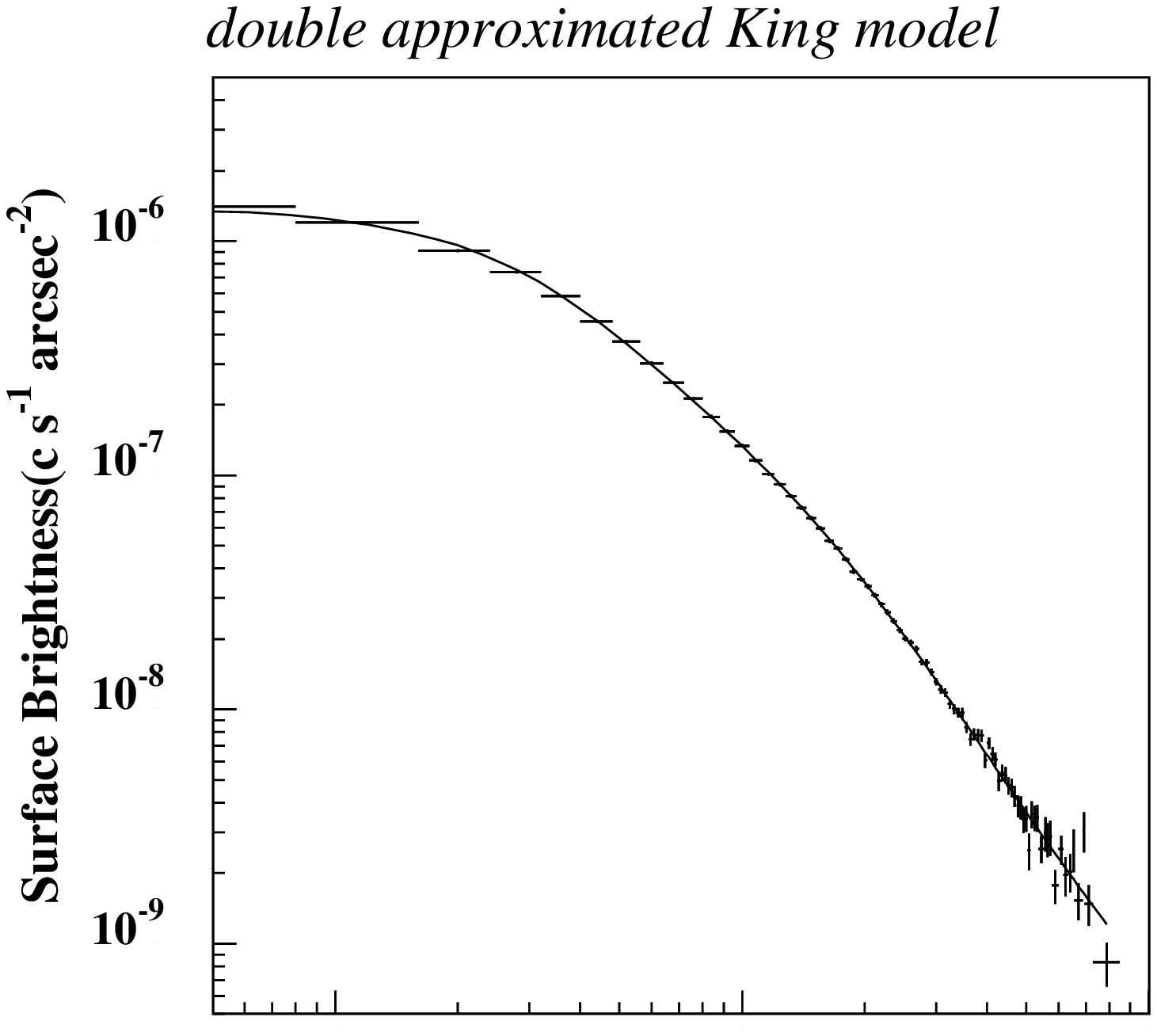,width=5.5cm}
\psfig{file=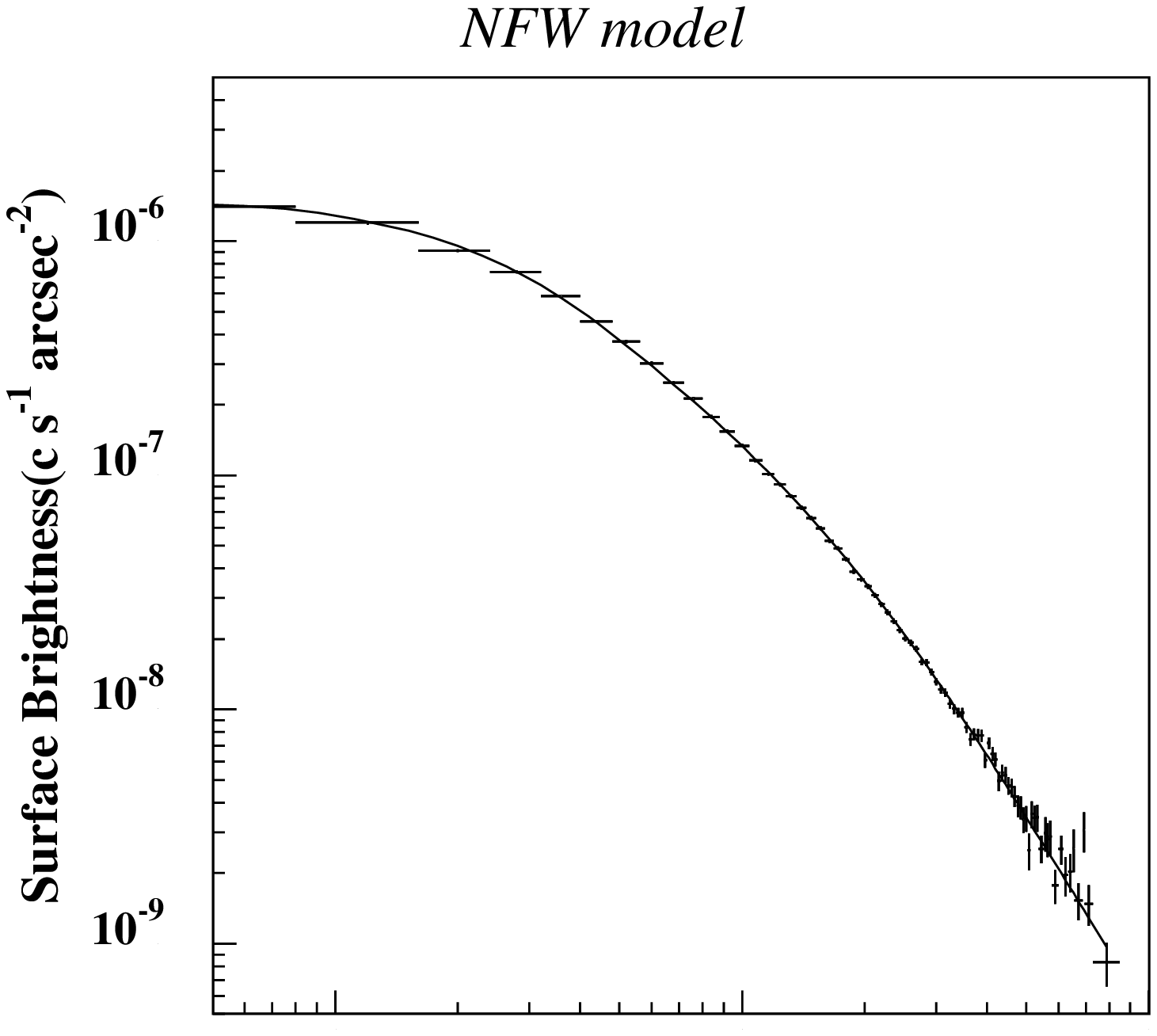,width=5.5cm}}
\vspace{-5mm}
\centerline{
\psfig{file=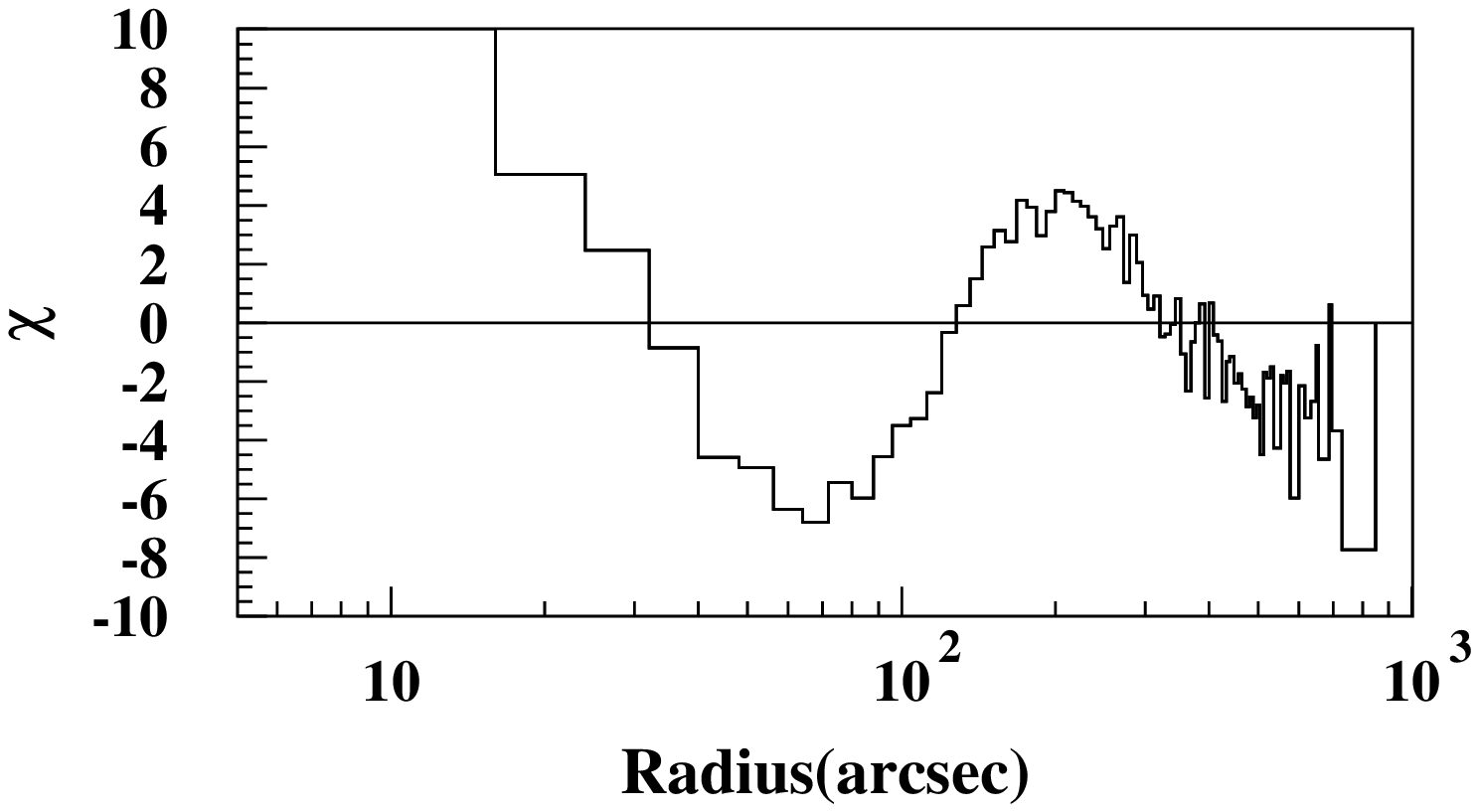,width=5.5cm}
\psfig{file=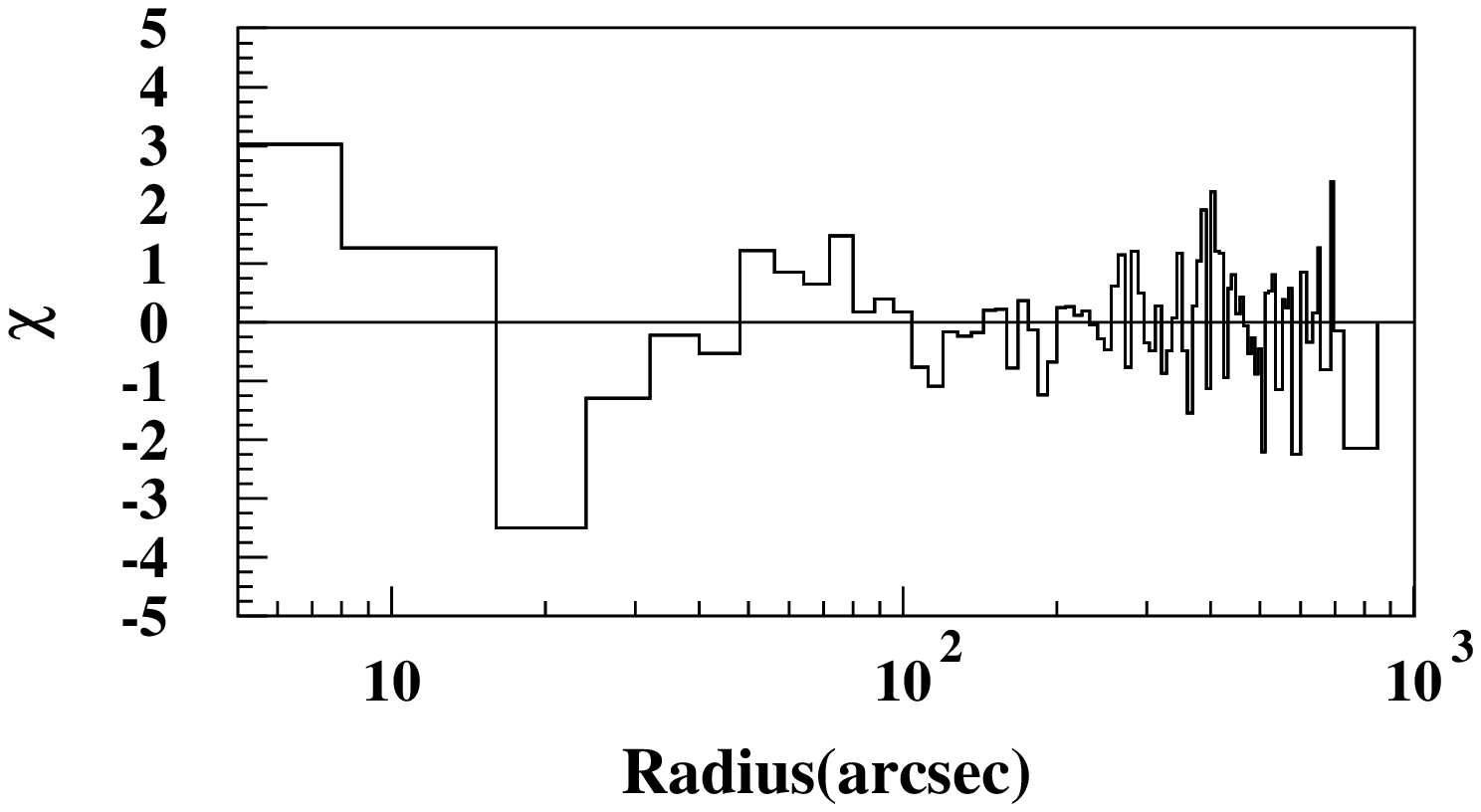,width=5.5cm}
\psfig{file=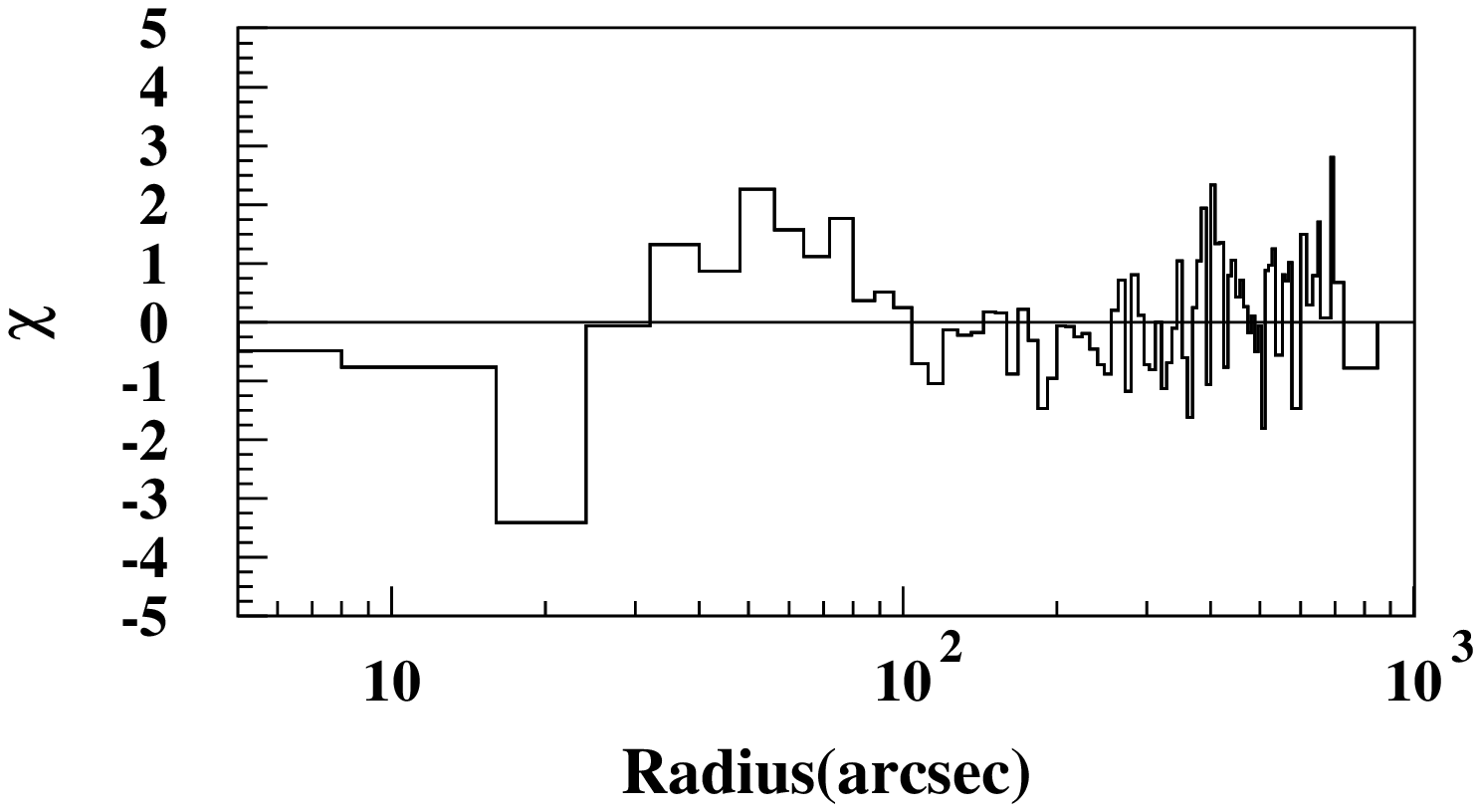,width=5.5cm}}
\caption{The upper panels show the 0.8--10~keV X-ray count rate profile
(crosses) and the best-fit model profiles
predicted from given total mass profiles.
The mass profile assumed in the model for the left, middle and right panel
is single approximated King model, double approximated King model
and NFW model, respectively.
The lower panels show residuals for the best-fit model profiles.
\label{fit3}}
\end{figure}

As for the mass profile,
we apply two representative models having a flat core and a cuspy core.
The King model (King 1966) is a classic model for the gravitational potential
structure characterized by a flat core.
We used an approximated formula of the King model,
in which the density profile is given as
\begin{equation}
\rho = \rho_0 \left[ 1 + \left(\frac{R}{R_{\rm c}}\right)^2 \right]^{-3/2} \ ,
\label{eq:king_rho}
\end{equation}
while the integrated mass profile is given as
\begin{equation}
M(<R) = 4\pi R_{\rm c}^3 \rho_0
\left[ \ln \left( \frac{R}{R_{\rm c}} + \sqrt{\frac{R^2}{R_{\rm c}^2} + 1 } \right)
     - \frac{R}{R_{\rm c}} \left(\frac{R^2}{R_{\rm c}^2} + 1 \right)^{-1/2} \right]\ ,
\label{eq:king_mass}
\end{equation}
where $\rho_0$ is the central density and $R_{\rm c}$ is the core radius
(see e.g. Binney and Tremaine 1987).
The predicted X-ray brightness profile from the King model
is fitted to the 0.8-10 keV count rate profile.
Fitting parameters are
the core radius ($R_{\rm c}$) and the central density ($\rho_0$) 
of the approximated King model, and the central ICM density ($n_0$).
The best-fit model and the fit residuals are shown in Fig.~\ref{fit3}
and the parameters derived are summarized in table~\ref{tab:mass_model}.
The fit is not acceptable.
In Fig.~\ref{masspro},
the best-fit King model is compared with the mass profile obtained
in Sect. 4.4,
showing clearly that a central mass excess is necessary
in additional to the King model mass.

\placefigure{fit3}

\placetable{tab:mass_model}

We then added another King-model component to account for
the central mass excess to construct a double approximated King model,
which still has a flat core profile.
Fitting parameters are now two sets with 
a core radius and a central density for each
($R_{\rm c,1}$, $\rho_{0,1}$, $R_{\rm c,2}$, and $\rho_{0,2}$),
and the central ICM density ($n_0$).
As shown in Fig.~\ref{fit3} and summarized in Table~\ref{tab:mass_model},
a good fit was obtained.

Another analytical formula for the total mass profile we employed
is the universal halo profile by Navarro et al. (1995, 1996, 1997; NFW model),
which is characterized by a sharp central cusp.
The density profile of the NFW model is given as
\begin{equation}
\rho = \rho_0 \left(\frac{R}{R_{\rm s}}\right)^{-1} 
              \left[ 1 + \left(\frac{R}{R_{\rm s}}\right)^2 \right]^{-2}\ ,
\end{equation}
where $R_{\rm s}$ is called scale radius.
The integrated mass profile of the NFW model is given as
(Makino et al. 1998; Suto et al. 1998)
\begin{equation}
M(<R) = 4\pi\rho_0 R_{\rm s}^3 
 \left[ \ln\left(1+\frac{R}{R_{\rm s}}\right) 
 - \frac{R}{R_{\rm s}}\left(1+\frac{R}{R_{\rm s}}\right)^{-1}\ \right]\ .
\end{equation}
The X-ray count rate profile predicted from the NFW model is fitted
to the data, where the free parameters are $\rho_0$, $R_{\rm s}$, and $n_0$.
As illustrated in Fig.~\ref{fit3} and summarized in Table~\ref{tab:mass_model},
the NFW model reproduces the data very well.

\begin{figure}
\epsscale{0.8}
\plotone{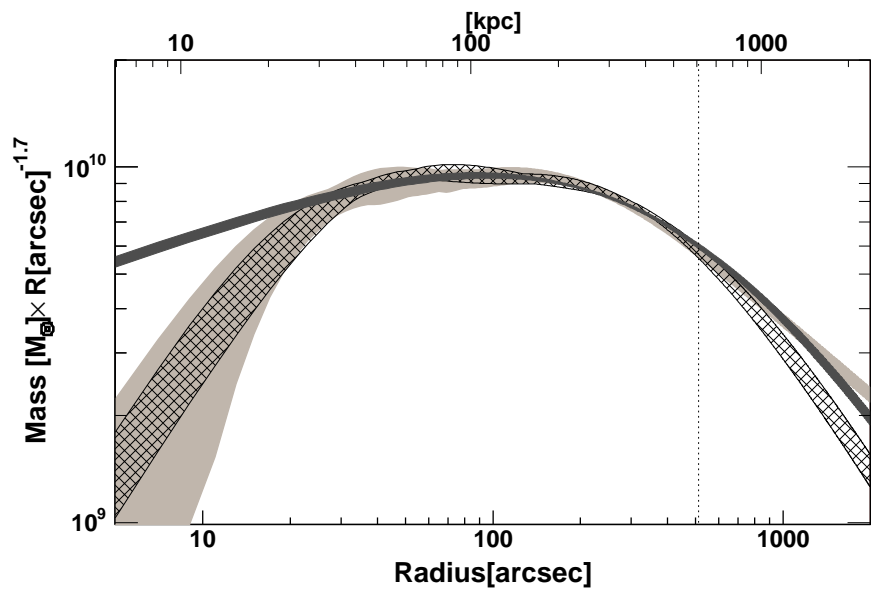}
\caption{The scaled total mass profiles obtained in Sect. 4.4 (light gray band)
and the two theoretical models,
the double approximated King model (hatched region with oblique lines)
and the NFW model (dark gray band) obtained in Sect. 4.5 are compared.
The vertical axis shows the integrated mass multiplied by $R^{-1.7}$.
The vertical dotted line is the same as in Fig.~\ref{masspro}.
\label{masspro_comp}}
\end{figure}

\begin{figure}
\epsscale{0.8}
\plotone{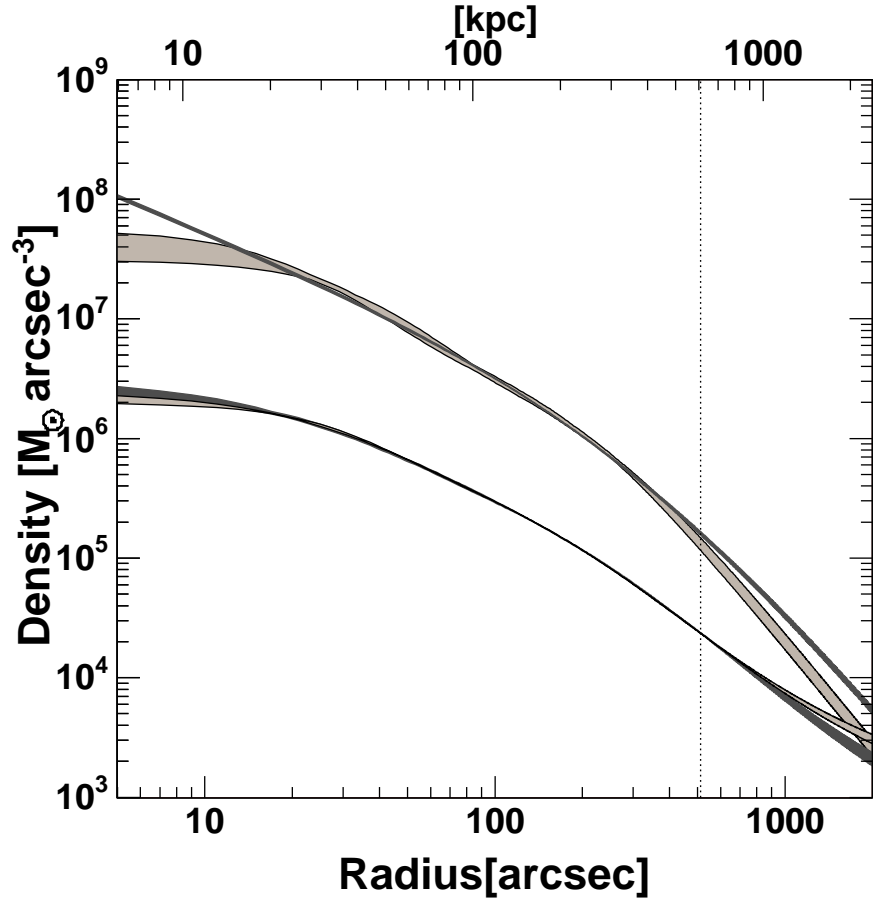}
\caption{The upper curves show the mass density profiles
of the total gravitating matter,
while the lower curves show the ICM density profiles.
The results with the double approximated King model and those with the NFW model
are illustrated with light gray band and dark gray band, respectively.
The vertical dotted line is the same as in Fig.~\ref{masspro}.
\label{dens_pro}}
\end{figure}

For each best-fit mass profiles,
the double approximated King model and the NFW model,
the corresponding virial mass, $M_{\rm vir}$,
and the virial radius, $R_{\rm vir}$, are calculated and given in Table~2.
The concentration parameters of the NFW model defined as
$c \equiv R_{\rm s}/R_{\rm vir}$ is obtained to be 6.1.
It is worth mentioning that the obtained $c$ and $M_{\rm vir}$
are consistent with the theoretically predicted $c-M_{\rm vir}$ relation
(e.g. Bullock et al. 2001).
The double approximated King model and the NFW model obtained above
are compared in Fig. \ref{masspro_comp},
together with the mass profile obtained in Sect. 4.4.
The three mass profiles are all consistent within
the 20-600 arcsec radius region,
and they approximately follow $M\propto R^{1.7}$ in this zone.
In Fig. \ref{dens_pro},
the double King model and the NFW model are compared
in the mass density profile,
together with the corresponding ICM density profile.
As clearly seen in Fig. \ref{masspro_comp} and \ref{dens_pro},
the two theoretical model profiles differ most significantly
in the very central region within $\sim$20 arcsec radius,
where XMM can not well resolve the spatial structure.
Ettori et al. (2002), using Chandra data,
obtained the total mass profile in A1795 with finer spatial resolution
and found that the density approximately follows $\rho \propto R^{-0.6}$
in the central region.
This profile is steeper than the King model ($\rho \propto R^0$)
and flatter than the NFW model ($\rho \propto R^{-1}$) near the center.
Therefore, the two mass profiles give a conservative mass range.
In the next subsection,
for determining the dark matter velocity dispersion profile,
we use both the double approximated King model and the NFW model
for the total mass profile model,
which would also be expected to give a conservative range of 
the velocity dispersion representing systematic uncertainty involved in
the usage of a specific mass profile model.

\placefigure{masspro_comp}

\placefigure{dens_pro}

\subsection{Dark matter velocity dispersion profile}

We now calculate the dark matter velocity dispersion profile
by solving the differential equation given in Eq. (3).
As the total mass profile, $M(<R)$,
the double approximated King model or the NFW model profile
obtained in Sect.~4.5 is employed,
while
the dark matter density profile is obtained from
the total mass and the corresponding ICM density profile by
$\rho_{\rm DM}(R) = 
\frac{1}{4\pi R^2}\frac{dM(<R)}{dR}-\mu m_{\rm p} n_{\rm g}(R)$.
Both $M(<R)$ and $\rho_{\rm DM}(R)$ are substituted in Eq. (\ref{eq:diffeq})
and the differential equation is solved with
the fourth-order Runge-Kutta method
for various boundary conditions at the center, i.e. $\sigma_{\rm DM}^2(0)$.

\begin{figure}
\epsscale{0.8}
\plotone{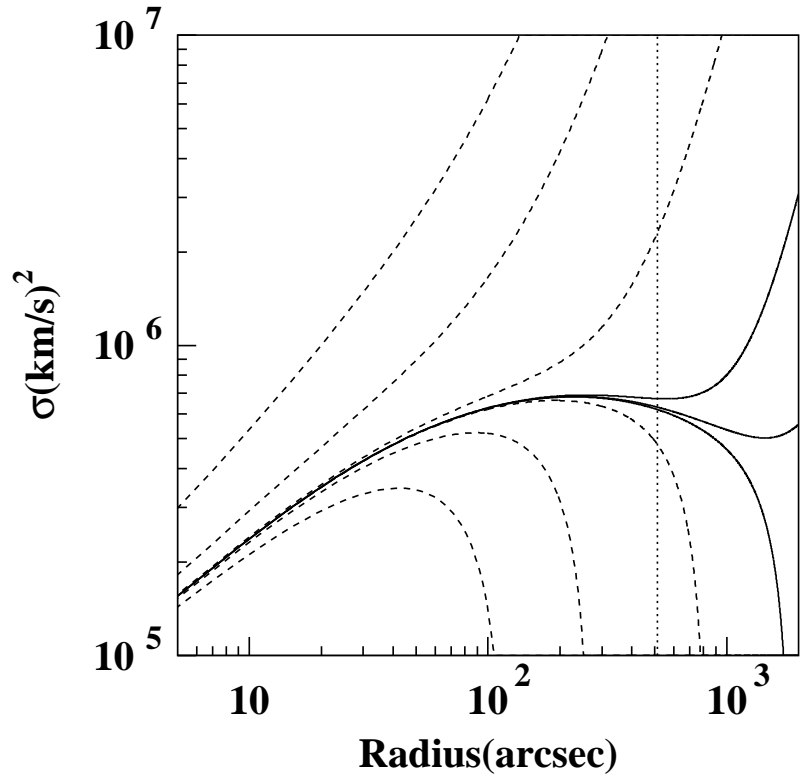}
\caption{Solutions of Eq. (3) with different $\sigma_{\rm DM}(0)$ values,
when the best-fit NFW model is used for the mass profile.
Physically plausible solutions are illustrated in the solid lines,
while the dashed lines are other possible solutions.
The vertical dotted line is the same as in Fig.~\ref{masspro}.
\label{setofsol}}
\end{figure}

Figure~\ref{setofsol} shows a set of solutions of Eq. (\ref{eq:diffeq}),
when the best-fit NFW model is applied to $M(<R)$.
Note that there is only a weak observational constraint
on the mass profile beyond $\sim1$ Mpc
and the solution at the large radii depends on the validity of the assumed
mass profile.
Depending on the inner boundary condition,
the dark matter velocity dispersion may fall to zero at small radii,
or it may diverge to infinity.
In order to select physically plausible solutions among them,
we placed a conservative boundary condition
that the velocity dispersion at the virial radius
is greater than 0 and less than the free fall velocity at this radius.

\placefigure{setofsol}

\begin{figure}
\epsscale{1.0}
\plotone{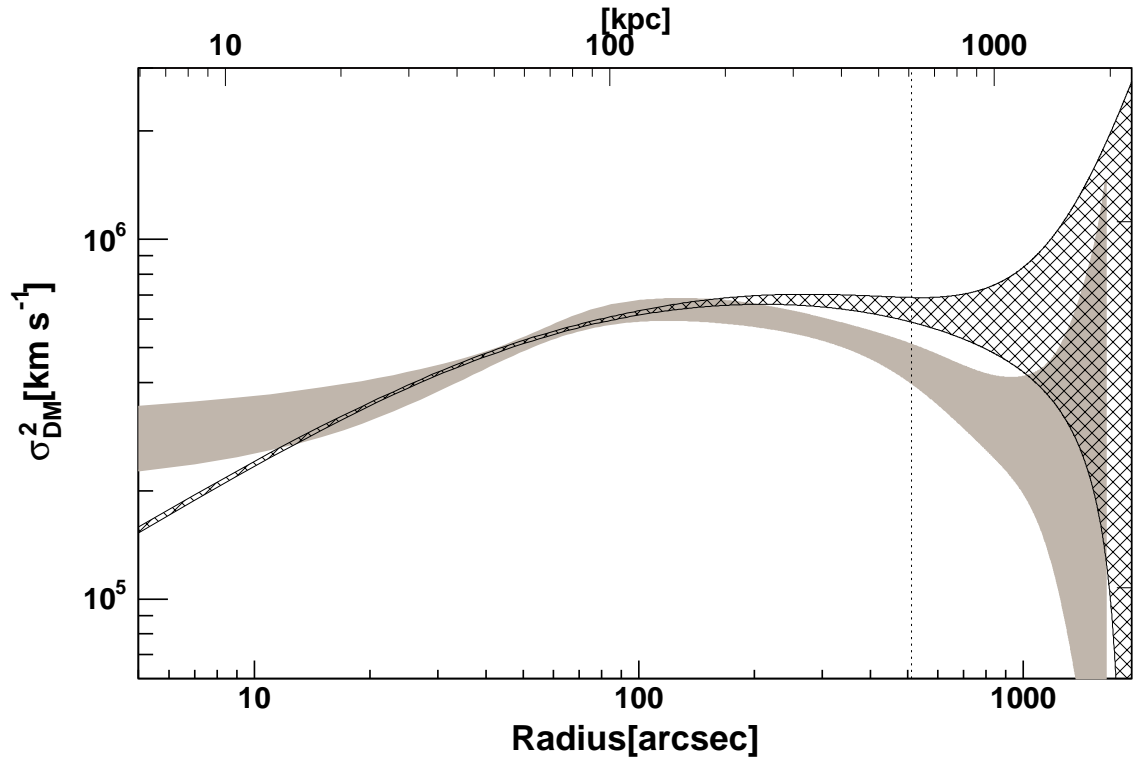}
\caption{The velocity dispersion profiles of the dark matter
are shown with the error bands.
The gray hatched region and the hatched region with oblique lines
are derived from the double approximated King model
and the NFW model as the total mass profile, respectively.
The vertical dotted line is the same as in Fig.~\ref{masspro}.
\label{sigma_dm}}
\end{figure}

\begin{figure}
\epsscale{1.0}
\plotone{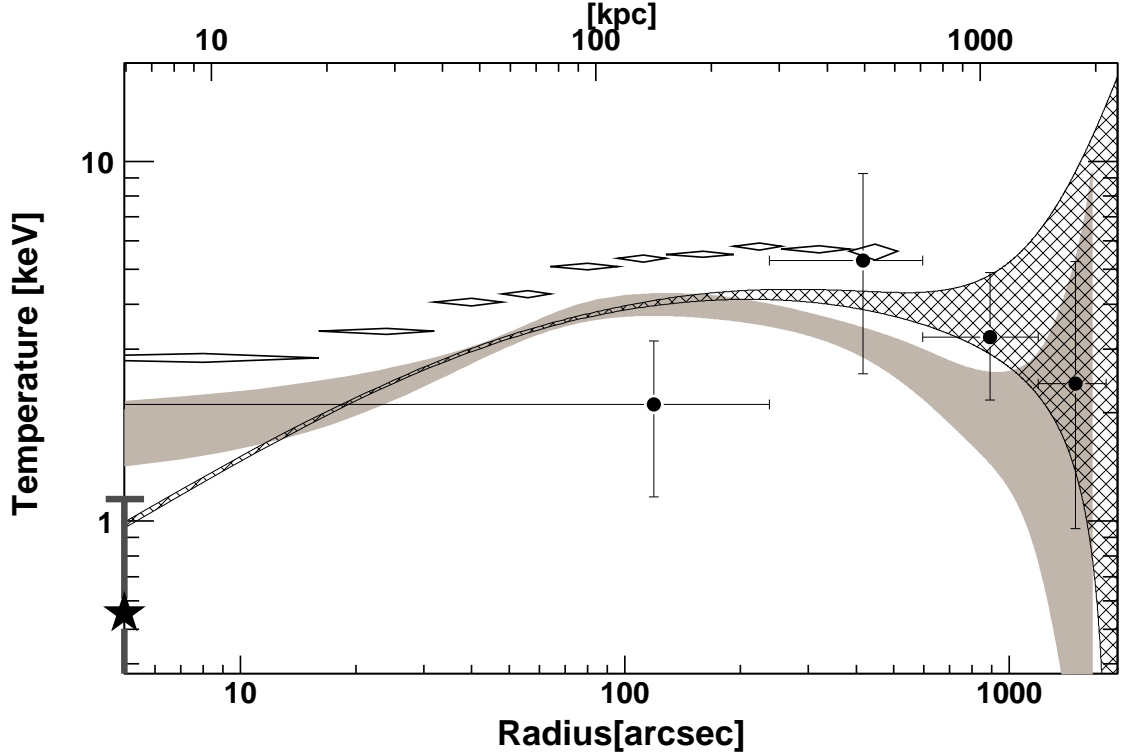}
\caption{The velocity dispersion profiles shown in Fig. \ref{sigma_dm}
are converted to ``temperature'' via
$kT_{\rm DM} = \mu m_{\rm p} \sigma_{\rm DM}^2$.
The ICM temperature profile derived in Sect. 4.3 and shown in Fig. \ref{tpro}
is overlaid with diamonds.
``Temperatures'' of member galaxies derived
by $\mu m_{\rm p} \sigma_{\rm gal}^2$
are also indicated with crosses,
where the galaxy velocity dispersions ($\sigma_{\rm gal}$) 
are taken from observations by den Hartog \& Katgert (1996).
The star symbol with the gray vertical error bar
indicates the central stellar velocity dispersion of the cD galaxy
being converted to ``temperature''
by $\mu m_{\rm p} \sigma_{\rm stellar}^2$,
where we use $\sigma_{\rm stellar} = 297\pm12$ (km s$^{-1}$)
as measured by Blakeslee \& Tonry (1992).
\label{dm_tpro}}
\end{figure}

\begin{figure}
\epsscale{1.0}
\plotone{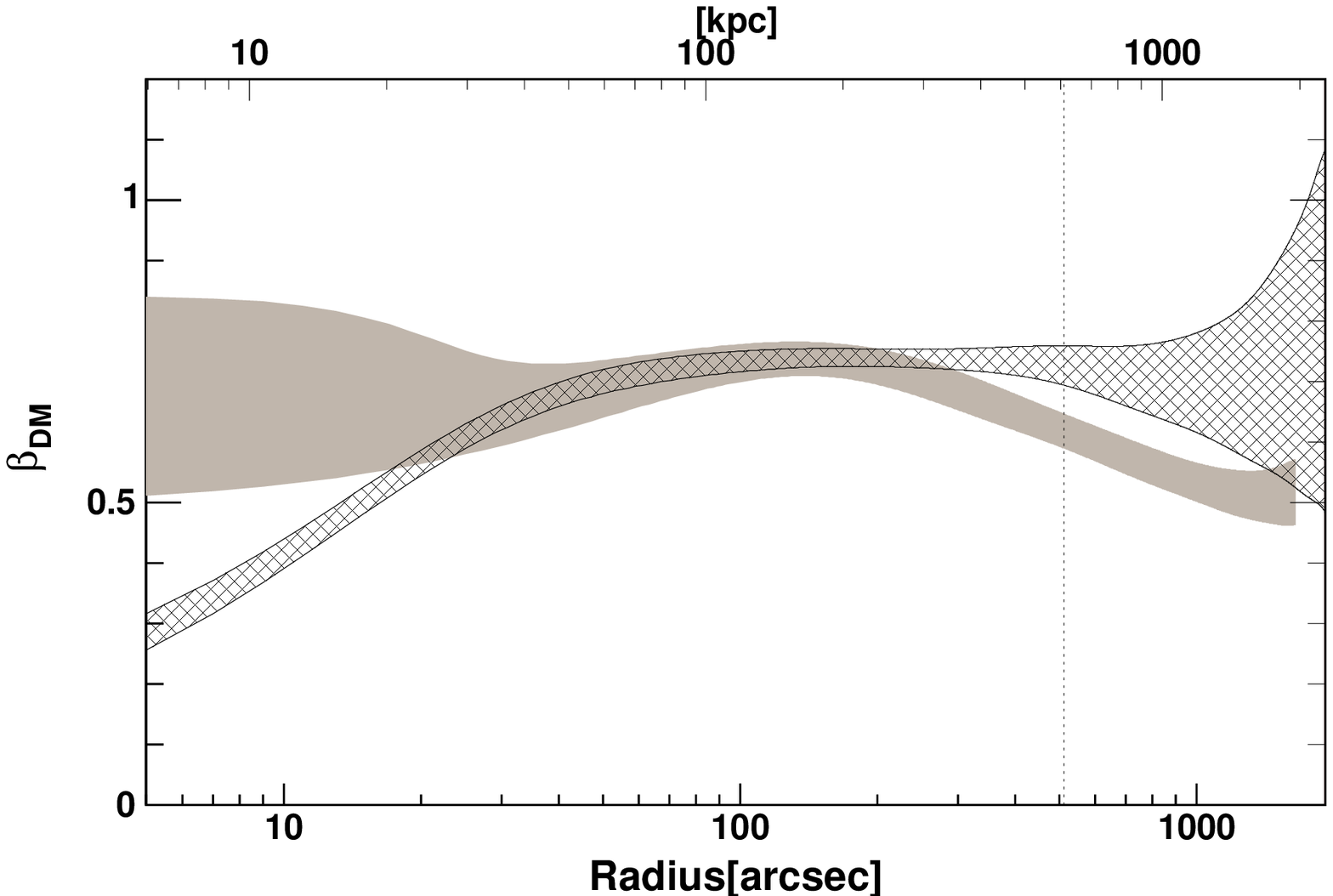}
\caption{Radial profile of $\beta_{\rm DM}$.
The gray region and the hatched region with oblique lines
correspond to the solutions shown in Fig.~\ref{sigma_dm}.
The vertical dotted line is the same as in Fig.~\ref{masspro}.
\label{beta_pro}}
\end{figure}

The constraints on the velocity dispersion profile thus obtained are shown
in Fig.~\ref{sigma_dm} individually for different mass models applied.
The errors on the ICM temperature profile are properly taken into
account to obtain the error bands.
In the central region, there is a clear difference
between the results from the double approximated King model
and the NFW model as expected from their different behavior.
The velocity dispersion is then converted to the ``temperature'' of
the dark matter ($k T_{\rm DM}\equiv\sigma_{\rm DM}^2 \mu m_{\rm p}$),
which is compared with the ICM temperature in Fig.~\ref{dm_tpro}.
The ICM temperature is greater than the dark matter ``temperature'' everywhere.
Even in the central region 
where radiative cooling is expected to be most effective,
the ICM temperature is significantly higher than that of the dark matter.
The comparison of the temperatures can be more directly described
by means of the $\beta_{\rm DM}$ value defined as 
$\beta_{\rm DM} \equiv \sigma_{\rm DM}^2 \mu m_{\rm p}/ k T = T_{\rm DM}/T$,
which ranges $\sim$0.3-0.8 (Fig.~\ref{beta_pro}).
There is no sign that the ICM is cooled significantly
below the dark matter ``temperature''.
In other words, the dark matter looks to form the temperature floor
that limits the ICM temperature.

As a matter of fact, what we derived by solving the Jeans equation
is the velocity dispersion of the collisionless particles
that includes the dark matter as well as the stellar component.
As shown in Fig.~\ref{masspro},
the stellar component makes a minor contribution to the total mass
and the derived velocity dispersion profile shown in Fig.~\ref{dm_tpro}
is virtually that of the dark matter,
except in the very central region, where the two components are
comparable.
If the galaxies alone are in a steady state,
they should also obey the Jeans equation under 
the same gravitational potential.
In Fig.~\ref{dm_tpro},
we overlay the velocity dispersion profile of the member galaxies measured
by den Hartog and Katgert (1996),
and the stellar velocity dispersion of the cD galaxy provided by
Blakeslee and Tonry (1992),
after converted to ``temperature'' via 
$k_{\rm B} T_{*} = \mu m_{\rm p} \sigma_{*}^2$.
They show roughly a consistent profile with that of the dark matter
at least in the case when the total mass profile is given by the NFW model.

\placefigure{sigma_dm}

\placefigure{dm_tpro}

\placefigure{beta_pro}

\section{Discussion}
\subsection{Anisotropy of $\sigma_{\rm DM}$}

Equation~(\ref{eq:Jeans}) that we have used to calculate the dark matter
velocity dispersion profile
is based on the assumption of isotropic motion of dark matter particles.
However, simulation studies
(e.g.  Eke, Navarro, and Frenk 1998; Col\'{i}n, Klypin, \& Kravtsov 2000)
indicate that the radial velocity dispersion, $\sigma_{\rm r}^2$,
should be rather larger than the tangential velocity dispersion,
$\sigma_{\rm t}^2 \equiv
\frac{1}{2}(\sigma_{\theta}^2+\sigma_{\phi}^2)$.
The degree of anisotropy is often measured with
\begin{equation}
A \equiv 1 - \frac{\sigma_{\rm t}^2}{\sigma_{\rm r}^2} \ ,
\label{eq:anisotropy}
\end{equation}
and
the Jeans equation is modified as
\begin{equation}
\frac{GM}{R} = 
- \sigma_{\rm DM}^2 \left( \frac{d \ln{\rho_{\rm DM}}}{d \ln{R}} 
 + \frac{d \ln{\sigma_{\rm DM}^2}}{d \ln{R}} + 2 A \right) \ .
\label{eq:Jeans2}
\end{equation}

\begin{figure}
\epsscale{1.0}
\plotone{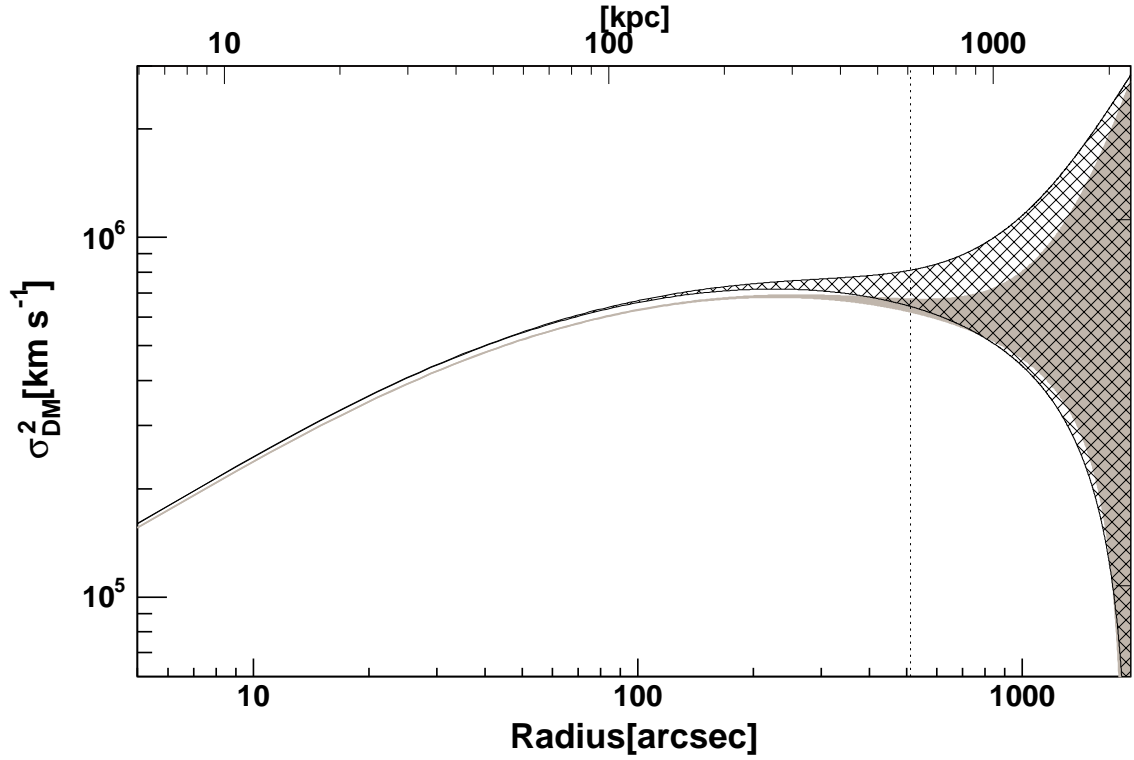}
\caption{The hatched region with oblique lines shows
the velocity dispersion profile of the dark matter,
when the NFW model is used as the total mass profile
and the anisotropy of the velocity distribution is introduced
as $A = 0.65 \frac{4 R/R_{\rm vir}}{(R/R_{\rm vir})^2 + 4}$.
The solution in the isotropic case is overlaid with the gray hatched region.
The vertical dotted line is the same as in Fig.~\ref{masspro}.
\label{sigma_dm2}}
\end{figure}

Employing $A = 0.65 \frac{4 R/R_{\rm vir}}{(R/R_{\rm vir})^2 + 4}$
derived by Col\'{i}n, Klypin, \& Kravtsov (2000),
which is as large as 0.5 at the virial radius, $R_{\rm vir}$, and
converging to 0 at the center,
we solved Eq. \ref{eq:Jeans2} as done in Sect. 4.6.
The $\sigma_{\rm DM}$ profile thus derived
for the case of the best fit NFW mass profile is shown in Fig.~\ref{sigma_dm2},
overlaid with the $A=0$ solution given in Fig.~\ref{sigma_dm}.
The $\sigma^2$ value with the anisotropy is greater than
that without the anisotropy by only $\sim$6\% at 100kpc
and at most $\sim$40\% around 1Mpc.

\placefigure{concern}

\placefigure{sigma_dm2}

\subsection{Heating source}

The $\beta_{\rm DM}$ profile determined here from observations
should provide information on the thermal history of the ICM.
From numerical simulation studies,
$\beta_{\rm DM}\sim 1-1.4$ is expected,
if there is no cooling or additional heating
(e.g. Metzler and Evrard 1994; Navarro et al. 1995; Bryan \& Norman 1998).
An obvious way to explain the $\beta_{\rm DM}$ value smaller
than unity in A1795 is heating of the ICM.

\begin{figure}
\epsscale{1.0}
\plotone{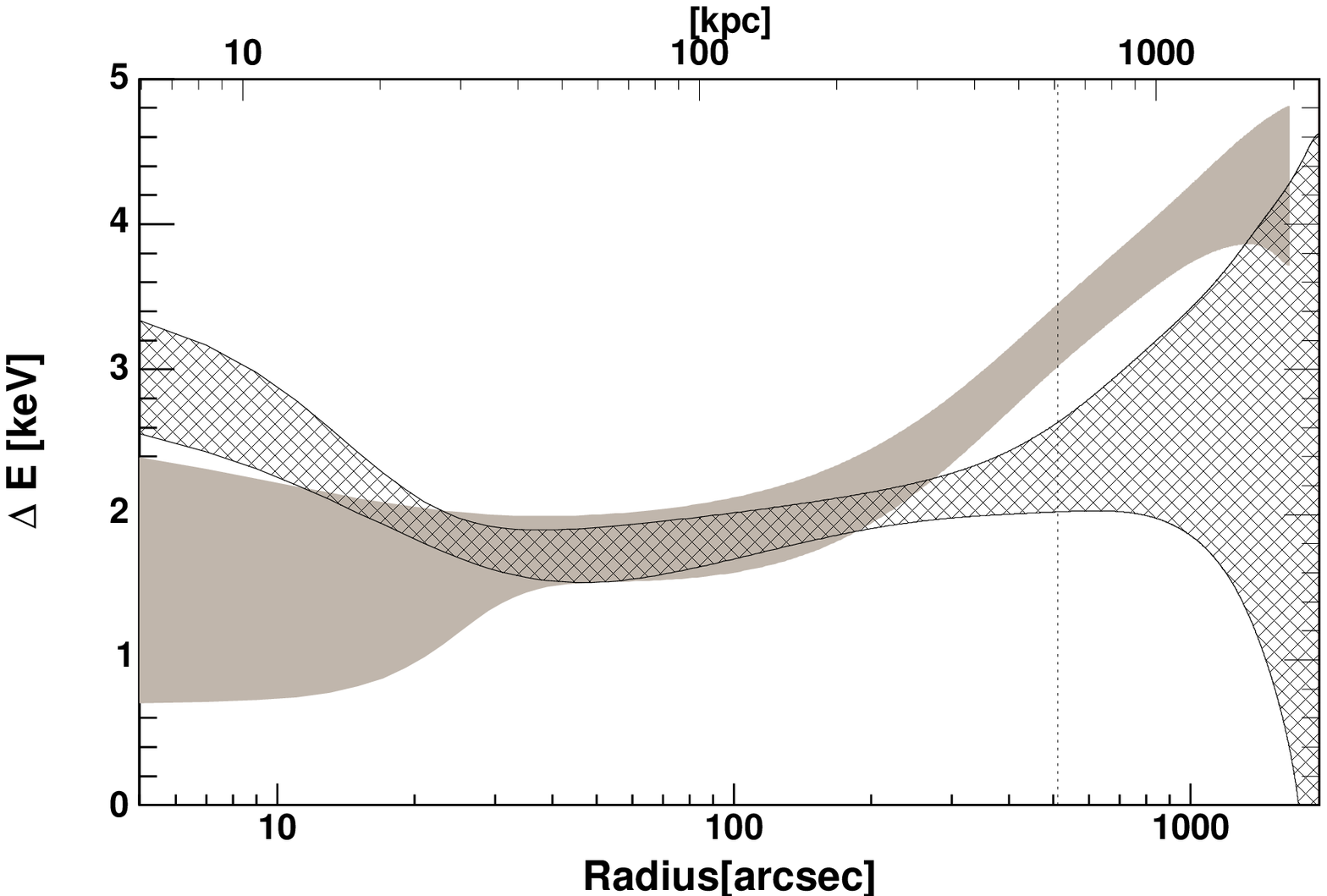}
\caption{Radial profile of $\Delta E$.
The gray region and the hatched region with oblique lines
correspond to the solutions shown in Fig.~\ref{sigma_dm}.
The vertical dotted line is the same as in Fig.~\ref{masspro}.
\label{mean_dE_pro}}
\end{figure}

\begin{figure}
\epsscale{1.0}
\plottwo{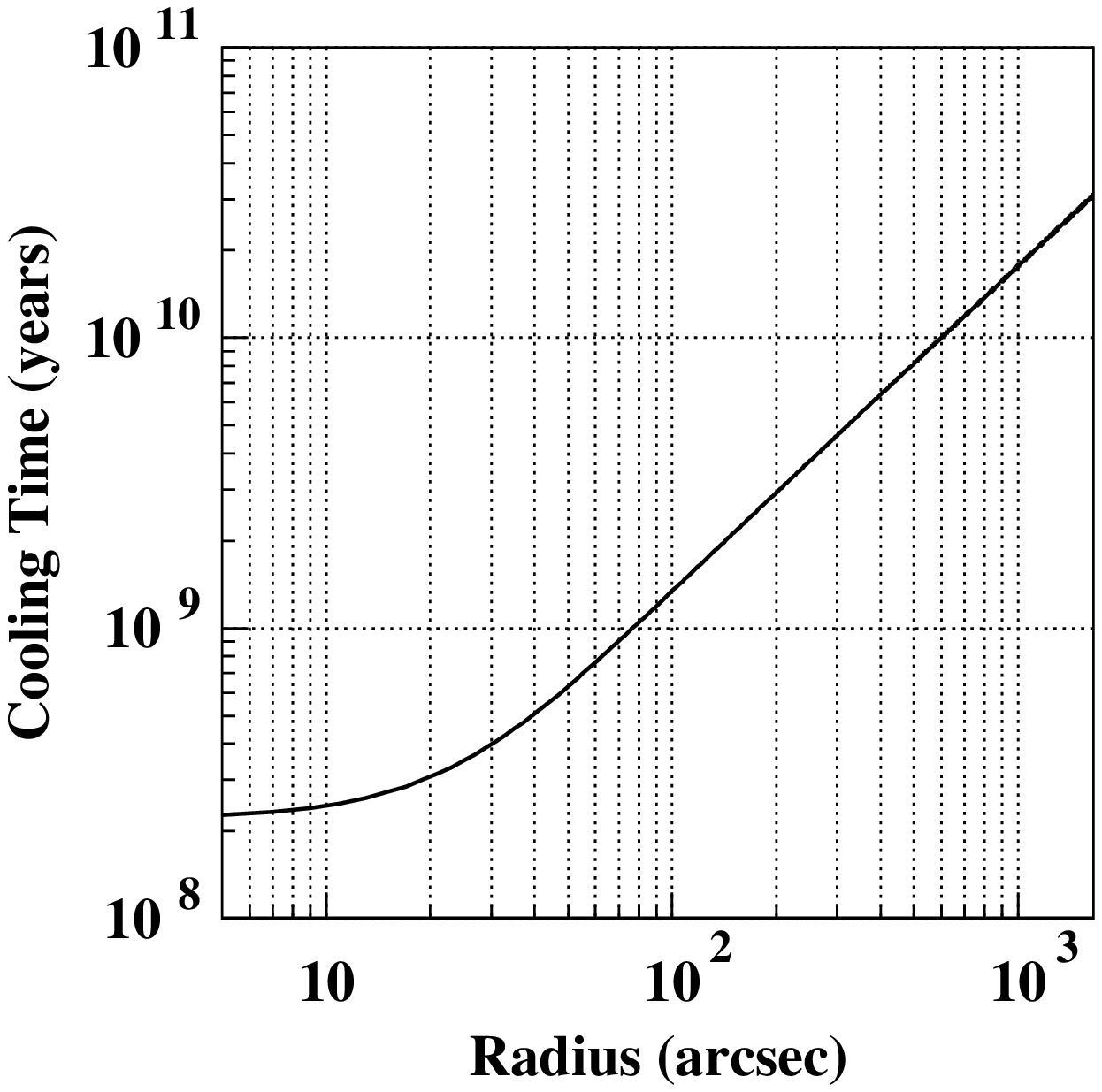}{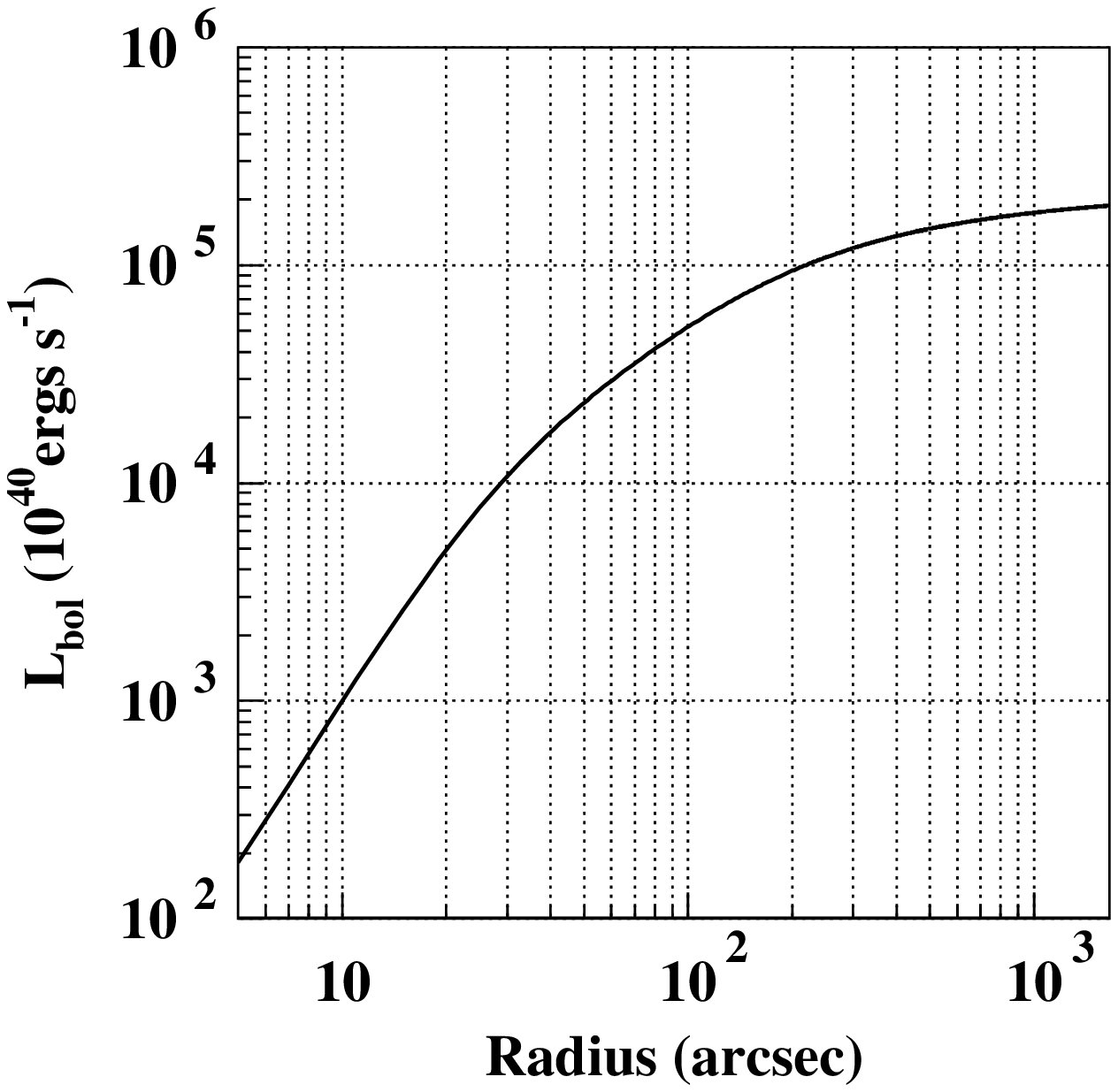}
\caption{(Left panel) Radiative cooling time of the ICM calculated
from the density and temperature at each radius.
(Right panel) Bolometric luminosity integrated within each radius.
\label{coolt}}
\end{figure}

As suggested from the break of the self-similarity between dark matter
and ICM,
there should have been non-gravitational heating acting globally.
We, from our results, estimated the excess energy of the ICM over that of
the dark matter particles as
\begin{equation}
\Delta E (<R) = \frac{3}{2}
\left( <kT> - \mu m_{\rm p} <\sigma_{\rm DM}^2> \right) \ ,
\end{equation}
where $<>$ denotes mass weighted mean within radius, $R$.
As shown in fig. \ref{mean_dE_pro},
the excess energy, $\Delta E$, thus derived
is found to be $\sim 1-3$ keV per particle,
which may be compared with theoretical model predictions.
The amount of energy injection to the gas phase that explains
e.g. the observed $L-T$ relation depends on the period
when the heating occurred.
Heating prior to cluster collapse, ``preheating'',
needs $0.1-0.3$~keV per particle
(e.g. Navarro et al. 1995; Tozzi \& Norman 2001),
while heating after a cluster formation requires
higher values of $1-3$~keV per particle
(e.g. Metzler \& Evrard 1994; Loewenstein 2000;
Wu, Fabian, \& Nulsen 2000; Bower et al. 2001).
Our results given above may indicate that
the global non-gravitational heating
that may cause the break of self similarity
has happened mainly within a collapsed cluster.
However, even if such non-gravitational heating that explains
the global X-ray feature of the cluster is provided,
the central region of A1795 has a short radiative cooling time
(Fig.~\ref{coolt}),
and there must be another significant energy input
to the ICM in the central region at the present epoch
to prevent the ICM from cooling.

\placefigure{mean_beta_pro}

\placefigure{mean_dE_pro}


As a possible energy source in the central region,
we first consider gravitational energy 
of the member galaxies
and stellar components therein.
The kinetic energy of the random motion of stars
can be partially transferred to the ICM by stellar mass loss.
Gas supplied by stellar mass loss has velocities of the bulk motions
of stars relative to the ICM, which is the sum of a galaxy motion
and the motions of the stars in the galaxy,
and is likely to be thermalized by interactions with the ambient gas.
If the stellar component moving in the same gravitational potential
has a similar velocity dispersion profile 
as that of the dark matter (Fig.~\ref{dm_tpro}),
the gas temperature achieved from this process is expected to
be comparable to the velocity dispersion of the dark matter.
This process nicely accounts for X-ray emission from
isolated X-ray compact elliptical galaxies (Matsushita 2001).
The input rate of the kinetic energy of the gas from stellar mass loss
may be simply estimated as
$\dot{E}=1/2 M_{\rm star} \dot{m} \sigma^2$ = $10^{42}$ ergs/s,
where $M_{\rm star}$ is the total stellar mass
of $1 \times 10^{12}$ M$_{\odot}$,
$\dot{m}$=$3\times10^{11}$ M$_{\odot}$ yr$^{-1}$ ($10^{11}$M$_{\odot}$)$^{-1}$
is the stellar mass loss rate in unit time,
and $\sigma$(=580$\pm$80 km/s) is the velocity dispersion
of the galaxies from den Hartog \& Katgert (1996; Fig. \ref{dm_tpro}).
(The stellar velocity dispersions in the galaxies are smaller
and neglected here.)
This is much smaller than the output energy in the central region
by X-ray radiation in galaxy clusters like A1795, however.
Thus for the case of galaxy clusters, we have to look for
additional heat sources.

Kinetic energy of the stellar component might be more efficiently
transferred to ICM via magnetic fields.
As pointed out by e.g. Makishima et al. (2001),
the motion of stars may amplify interstellar magnetic fields
and reconnections of the fields may heat up the ICM rather efficiently.
The galaxies must have lost their kinetic energies through interactions
with the ICM and have gradually fallen inwards accumulating
onto the central galaxy to form the cD galaxy.
A deep optical image of the cD galaxy in A1795 derived
by Johnstone et al. (1991) shows a concentration of elliptical galaxies
of various sizes and stars forming a largely extended
envelope with 131 $h_{70}^{-1}$~kpc effective radius,
which strongly suggests the on-going formation process of the cD galaxy.
Quantitatively, the total amount of dynamical energy of 
the stellar component in the member galaxies
that has been lost in the past is estimated.
The stellar component in the galaxies
is assumed to have a negligible small potential ($U$)
and kinetic energy ($K$) before the formation of the cluster, 
and the current energy of the stars 
is estimated to be $U+K\sim - 10^{62}$ ergs.
If the energy has been released over the last 10 Gyr
and has been spent in ICM heating,
the heating luminosity is expected to be $\sim3\times10^{44}$ ergs s$^{-1}$.
This amounts to the bolometric luminosity of the ICM
within 60$h_{70}^{-1}$~kpc (Fig. \ref{coolt}), and
may be sufficient to sustain the thermal energy of the ICM
against radiative cooling.
This model predicts that the stellar velocity dispersion became
smaller than that of ICM, i.e. $\beta_{\rm spec} < 1$,
which is consistent with the actual observed value
in the central region (Fig. \ref{dm_tpro}).

Alternatively, there may be sufficient non-gravitational heat input
from an active galactic nucleus (AGN) powered by 
an accretion of low entropy gas at a cluster center
(Churazov et al. 2002; B\"{o}hringer et al. 2002).
Numerical simulations show that
an outflow from an AGN may form hot bubbles of relativistic plasma
rising with buoyancy, and
the bubbles may uplift cold gas mixing with ambient ICM
(Churazov et al. 2001; Quilis, Bower, \& Balogh 2001;
Br\"{u}ggen \& Kaiser 2002; Basson \& Alexander 2003).
High resolution imaging observations revealed 
ripples and shock features in the central region
of Perseus cluster (Fabian et al. 2003) and M87 (Forman et al. 2004),
suggesting
that the bubble energy may also be transfered by sound wave to larger
distances,
and that shocks may be the major contribution of the energy dissipation.
The heating mechanism is self regulated:
the lower the entropy, the higher the accretion rate onto the central engine.
A portion of the accretion power is dissipated back into the ICM
to make its entropy high and regulate the accretion rate to achieve
an equilibrium state.
This process automatically prevents the persistence of cold
and hence dense clouds.
The cD galaxy of A1795 has a radio source, 4C26.42,
and the existence of an AGN is evident.
Using the physical state of the ICM in the center
we can actually estimate the energy provided by the AGN
by applying the Bondi accretion model.
According to the well-known correlation of the black hole mass
with the mass of the bulge component (Magorrian et al. 1998),
the black hole mass is expected to be $\sim6\times 10^9$ M$_{\odot}$.
Assuming that the gas profile is flat in the center,
we can use the measured values of $n_{\rm g}$=0.1 cm$^{-3}$
and $T_0$=2.8 keV to obtain the Bondi mass accretion rate
\begin{eqnarray}
\dot{M} & = & 4\pi 0.25 \rho_{\infty} c_{s,\infty}^{-3} (G M_{\rm BH})^2 \\
        & = & 0.23 M_{\odot}/{\rm yr}
   \left( \frac{n}{1 {\rm cm}^{-3}} \right)
   \left( \frac{T}{1 {\rm keV}} \right)^{-3/2}
   \left( \frac{M_{\rm BH}}{6\times 10^9 M_{\odot}} \right)^2 \ ,
\end{eqnarray}
where $\rho_{\infty}$ and $c_{s,\infty}$ are the density and sound velocity
outside the Bondi accretion radius.
We find $\dot{M} \sim 0.4 \times 10^{-2} M_{\odot}/{\rm yr}$.
Under the standard assumption of 10\% of the accretion energy 
to be dissipated,
the output energy is found to be 
$E = 0.1 \dot{M} c^2 \sim 3 \times 10^{43}$ ergs/s.
This amounts to the X-ray luminosity within the inner $\sim20$ kpc
region only.
The Bondi accretion radius is estimated to be 
$R_{\rm B} \sim G M_{\rm BH} / c_{\rm s}^2 \sim$ 30 pc,
much smaller than the resolution of
the temperature and density structure that can be measured with XMM-Newton.
If the ICM density is not uniform but is clumpy,
the Bondi accretion rate should be significantly larger
and the heating rate could also be larger than the above estimation.
We can note that the ICM conditions might temporally vary
and that we currently see a relatively low state.

Note that there are other heating mechanisms discussed,
which include e.g. a classical idea of thermal conduction
from the hot outer regions
(e.g. Takahara \& Takahara 1979; Tucker \& Rosner 1983;
Bregman, \& David 1988; Gaetz 1989),
and the acoustic wave heating 
recently proposed by Fujita, Suzuki, \& Wada (2004).
Comparison of those model predictions with our results presented
in the current paper would be very interesting.

\placefigure{coolt}

\section{Summary}
We derived the dark matter velocity dispersion profile
from an X-ray observation for the first time.
Using the XMM-Newton EPIC data of A1795,
we derived $\beta_{\rm DM}$ of $\sim0.3-0.8$,
indicating that the ICM temperature is larger than the dark matter
``temperature'' everywhere.
We also derived the excess energy in the ICM, $\Delta E$,
which is found to be $\sim1-3$~keV per particle.
These observational quantity should be useful 
to provide a new clue to solve the heating mechanism 
in the cooling core
and to understand the thermal history of a galaxy cluster.


\acknowledgments
We thank the XMM team for providing the data and the data analysis tools.
In particular, we express our gratitude to Michael Freyberg for
helping the data analysis.
We also acknowledge FTOOLS.
We thank Kuniaki Masai for helpful discussions.
We are also grateful to Paul Lynum for critically reading this manuscript.

\clearpage


\clearpage 

\clearpage

\begin{deluxetable}{lll}
\tablecaption{Best-fit parameters of the double $\beta$ model
fitted to the 0.8-10 keV brightness profile
\label{tab:2beta}}
\tablewidth{0pt}
\tablehead{
\colhead{parameter} & \colhead{value$^{a)}$}   & \colhead{}
}
\startdata
$\Sigma_{0,1}$ & $(1.42\pm0.04)\times10^{-6}$  & [c/s/arcsec$^2$] \\
$R_{\rm c,1}$  & 26.3  $\pm$ 1.0                & [arcsec] \\
$\beta_1$      & 0.562 $\pm$ 0.015            &  \\
$\Sigma_{0,2}$ & $(1.92\pm0.12)\times10^{-7}$ & [c/s/arcsec$^2$] \\
$R_{\rm c,2}$  & 106    $\pm$ 4                 & [arcsec] \\
$\beta_2$      & 0.633  $\pm$ 0.010             &  \\
$\chi^2/\nu$   & 83.2/74                        &  \\
\enddata
\tablenotetext{a}{Errors are 90\% ($\Delta \chi^2 = 2.7$) confidence.}
\end{deluxetable}

\begin{deluxetable}{ccccccccc}
\tabletypesize{\scriptsize}
\tablecaption{\label{tab:mass_model}}
\tablewidth{0pt}
\tablehead{
\colhead{mass}           &  \colhead{$\rho_{0,1}$}   &
\colhead{$R_{\rm c,1}$}  &  \colhead{$\rho_{0,2}$}   &
\colhead{$R_{\rm c,2}$}  &  \colhead{$n_{\rm g}(0)$} &
\colhead{$\chi^2/\nu$}   &  \colhead{$R_{\rm vir}^{a)}$}  &
\colhead{$M_{\rm vir}^{a)}$}  \\

\colhead{model}     & \colhead{[ M$_{\odot}$arcsec$^{-3}$ ]} &
\colhead{[arcsec]}       & \colhead{[ M$_{\odot}$arcsec$^{-3}$ ]} &
\colhead{[arcsec]}       & \colhead{[cm$^{-3}$]}     &
\colhead{}               & \colhead{[arcsec (Mpc)]}             &
\colhead{[$10^{14}$M$_{\odot}$]}

}
\startdata
1 King  & $1.28\times10^{7}$ & 94.6 & --  & -- & 0.063 & 1124/77 & - & -\\
2 King  & $3.31^{+1.76}_{-0.42}\times10^{7}$ & $34.8^{+6.9}_{-10.9}$ & $2.74^{+1.54}_{-0.79}\times10^{6}$ & $194.4^{+26.4}_{-34.7}$ & $0.086^{+0.014}_{-0.006}$
& 88.3/75 & $1716^{+63}_{-74}$ ($2.04^{+0.07}_{-0.09}$) & $5.38^{+0.62}_{-0.66}$ \\
\tableline
             &  $\rho_0$  &  $R_{\rm s}$  &  &  &  & &  & \\
NFW          & $1.64^{+0.18}_{-0.13}\times10^{6}$ & $330.2^{+16.1}_{-21.9}$ & -- & -- & $0.118^{+0.022}_{-0.012}$
& 88.7/77 & $1958^{+26}_{-44}$ ($2.33^{+0.03}_{-0.06}$) & $8.01^{+0.32}_{-0.54}$ \\
\enddata
\tablenotetext{a}{
$M_{\rm vir} \equiv \frac{4}{3}\pi R_{\rm vir}^3 \rho_0 (1+z)^3
\Delta$,
where $\rho_0$ is the mean density of the universe at present,
and $\Delta$ is the ratio between a cluster mean density and the mean
density of the universe at the cluster-formation redshift $z$ 
(=0.0616 for A1795) (see e.g. Kitayama \& Suto 1996).
}
\end{deluxetable}

\end{document}